\documentclass{article}
\usepackage[utf8]{inputenc}
\usepackage{graphicx}
\usepackage{tabularx}
\usepackage{natbib}
\usepackage{float}
\usepackage{caption}
\usepackage{subcaption}
\usepackage{amsmath}
\usepackage{graphicx} 
\usepackage{booktabs}
\usepackage{apalike}
\usepackage{amsmath,amsfonts,amssymb,amsthm}
\usepackage[dvipsnames]{xcolor}
\usepackage{titling}
\usepackage{lipsum}
\usepackage{multirow}
\usepackage{tabu}
\usepackage{mathtools}
\usepackage[thinlines]{easytable}
\usepackage{array}
\usepackage[title, titletoc, toc]{appendix} 
\usepackage{afterpage} 

\usepackage{algorithm}
\usepackage{algorithmic}
\usepackage{tikz}
\usepackage{bm}
\usepackage{color, colortbl}
\usepackage{hhline}
\usepackage{makecell} 
\usetikzlibrary{arrows,matrix,positioning,fit, decorations.pathreplacing}
\usepackage{titlesec}
\titleformat*{\section}{\large\bfseries}
\titleformat*{\subsection}{\normalsize\bfseries}
\titleformat*{\subsubsection}{\normalsize\bfseries}
\titleformat*{\paragraph}{\normalsize\bfseries}
\titleformat*{\subparagraph}{\normalsize\bfseries}
\usepackage{setspace}
\usepackage{longtable}

\usepackage[left=1in,top=1in,right=1in,bottom=1in,nohead,paperwidth=8.5in, paperheight=11in]{geometry}
\usepackage{authblk}
\usepackage{xurl}
\theoremstyle{definition}
 
\usepackage{url}

\usepackage[normalem]{ulem}
\usepackage{color}
\newcolumntype{Y}{>{\centering\arraybackslash}X}

\begin{document}
\title{\bf 
Hierarchical Regularizers for Reverse Unrestricted Mixed Data Sampling Regressions}
\author{Alain Hecq\thanks{a.hecq@maastrichtuniversity.nl} \quad Marie Ternes\thanks{m.ternes@maastrichtuniversity.nl} \quad Ines Wilms\thanks{Corresponding author: Ines Wilms, Maastricht University, School of Business and Economics, P.O. Box 616, 6200 MD Maastricht, The Netherlands, Email: i.wilms@maastrichtuniversity.nl}
}
\affil{Department of Quantitative Economics, Maastricht University}
\date{\today}

\maketitle
\begin{abstract}
Reverse Unrestricted MIxed DAta Sampling (RU-MIDAS) regressions are used to model high-frequency responses by means of low-frequency variables. However, due to the periodic structure of RU-MIDAS regressions, the dimensionality grows quickly if the frequency mismatch between the high- and low-frequency variables is large. Additionally the number of high-frequency observations available for estimation decreases. 
We propose to counteract this reduction in sample size by pooling the high-frequency coefficients and further reduce the dimensionality through a sparsity-inducing convex regularizer that accounts for the temporal ordering among the different lags. To this end, the regularizer prioritizes the inclusion of lagged coefficients according to the recency of the information they contain. We demonstrate the proposed method on 
two empirical applications, one on realized volatility forecasting with macroeconomic data and another on demand forecasting for a bicycle-sharing system with ridership data on other transportation types.
\end{abstract}

\paragraph{Acknowledgements.} 
The third author was financially supported by the Dutch Research Council (NWO) under grant number VI.Vidi.211.032. In addition, we acknowledge support through the HiTEc Cost Action CA21163. We gratefully acknowledge the comments by participants at the 16th International Conference on Computational and Financial Econometrics (CFE 2022) and the 9th Annual Conference of the International Association for Applied Econometrics (IAAE 2023) and thank Claudia Foroni and Stephan Smeekes for helpful discussions.

\newpage
\section{Introduction}
In recent years, a growing body of literature on mixed-frequency (MF) models arose to exploit the information available in series recorded at different frequencies. So far, the literature has mostly focused on modeling low-frequency (LF) variables by means of more timely, high-frequency (HF) variables to improve the former's forecast or nowcast accuracy. One commonly used univariate MF model is the MIxed DAta Sampling (MIDAS) regression (\citealp{ghysels2004midas}) that links low-frequency to high-frequency data through tightly restricted distributed lag polynomials. Various extensions to the MIDAS framework have been proposed in the literature. \cite{foroni2015umidas}, for instance, consider univariate unrestricted MIDAS (U-MIDAS) to enable ordinary least squares estimation instead of non-linear least squares (NLS). \cite{ghysels2016macroeconomics} propose a multivariate extension to model relations between high- and low-frequency series in a stacked mixed-frequency VAR (MF-VAR) system; state-space alternatives of the latter are provided by, amongst others, \cite{mariano2003new}, \cite{schorfheide2015real} and \cite{koelbl2020new}. Recently, high-dimensional MF models have been proposed for the univariate (e.g., \citealp{han2020high}; \citealp{mogliani2021bayesian}; \citealp{babii2022machine}), multivariate (for factor models see e.g., \citealp{marcellino2010factor}; \citealp{foroni2014comparison}; \citealp{andreou2019inference}; for Bayesian estimation e.g., \citealp{schorfheide2015real};  \citealp{ghysels2016macroeconomics}; \citealp{gotz2016testing}; \citealp{mccracken2020real}; \citealp{cimadomo2021nowcasting}; \citealp{paccagnini2021identifying} and for sparsity-inducing regularizers  e.g., \citealp{hecq2022hierarchical}) and panel settings (e.g., \citealp{babii2022panel}). 

A less explored research area concerns MF models suited for modeling HF variables by means of LF variables. This can be of interest when fundamental economic variables, usually available at a lower frequency, play an important role for forecasting higher frequency variables as they contain more valuable information for longer-term horizons. While the stacked MF-VAR of \cite{ghysels2016macroeconomics} can be used to predict HF variables using LF ones (despite its main usage for LF variables)\footnote{Another example is \cite{dalBianco2012short} who model the euro-dollar exchange rate at weekly frequency exploiting a set of monthly financial and macroeconomic variables using a state-space framework as in \cite{mariano2003new}.}, dedicated MF models for HF components have been proposed in the literature. \cite{foroni2018rmidas} extend the univariate MIDAS and U-MIDAS models to the Reverse (R-MIDAS) and Reverse Unrestricted MIDAS (RU-MIDAS) model where the HF dependent variable is regressed on the LF explanatory variables. Indeed, in RU-MIDAS models each HF component has its own model equation, thereby allowing the (lagged) dynamics between the high and low frequency variables to change for every HF component. We consider this RU-MIDAS model as it can be more flexible and parsimonious than the multivariate stacked MF-VAR and does not require the specification of a model for the LF variable. Moreover, the RU-MIDAS can automatically incorporate HF information released within an ongoing LF period (i.e., a conditional model), whereas the MF-VAR would require one to use a structural MF-VAR that describes the contemporaneous correlations among variables to update the forecast for every HF period. Thus, the RU-MIDAS model is more appropriate for direct forecasts of the HF variable than the MF-VAR.

Previous studies have, however, mostly considered RU-MIDAS models where the frequency mismatch $m$ between the HF and LF variable is small, for instance a setting with monthly HF components and a quarterly LF component (see e.g., \citealp{foroni2018rmidas}). Our interest lies in settings where this frequency mismatch $m$ is large,   think for instance of a setting with daily HF components and a monthly LF component; such settings only recently attracted attention (see e.g., \citealp{foroni2023low}). In this case RU-MIDAS regressions are, however, severely affected by the ``curse of dimensionality". There are two main sources of this curse. First, the larger the frequency mismatch the more equations, namely one for each HF component, need to be estimated. Since each equation comes with its own separate set of parameters, this in turn increases the total number of parameters in the RU-MIDAS model. Second, the number of observations available for the estimation of each HF equation is limited to the number of LF observations. Hence, the larger $m$, the fewer observations are available for estimation since the sample size of the high-frequency variable is divided by $m$. Without further adjustments, one would be limited to use RU-MIDAS with a small frequency mismatch. In this paper, we address this curse of dimensionality by proposing a pooled RU-MIDAS model with a hierarchical convex regularizer. First, to address the issue of sample size reduction, we propose to pool the regression coefficients regulating the lagged HF dynamics across the $m$ different HF equations. Pooling relies on the assumption that the lagged effect of a HF variable on itself does not change with the HF component that is modelled. For instance, by pooling we constrain the lagged effect of the previous day to be the same for each day within a month. Similarly for the lagged effect of order two and so on. We advocate to pool in RU-MIDAS models where the frequency mismatch $m$ is large as it allows one to use the full HF sample size instead of the limited LF sample size to estimate the HF coefficients, which is a considerable difference when the frequency mismatch is large. As a by-product, by pooling we decrease the total number of parameters that need to be estimated.

Second, to further reduce the curse of dimensionality in the pooled RU-MIDAS model one could, for instance, turn to distributed lag polynomials that restrict the coefficients similar to restricted R-MIDAS models. However, R-MIDAS (similar to the regular MIDAS) requires computationally expensive NLS estimation. Alternatively, \cite{foroni2023low}  consider RU-MIDAS in a Bayesian framework, whose good performance has been demonstrated for forecasting electricity prices \citep{foroni2023low} and shipping freight-cost indices \citep{bouri2022role}. We instead consider sparsity-inducing regularizers which form an appealing alternative (see \citealp{hastie2015statistical} for an introduction) and have successfully been applied to regular MIDAS regressions (\citealp{babii2022machine}) and  MF-VARs (\citealp{hecq2022hierarchical}). However, to the best of our knowledge they have not been explored yet as tool for dimension reduction in RU-MIDAS regressions. In this paper, we adapt the mixed-frequency hierarchical regularizer of \cite{hecq2022hierarchical} to the RU-MIDAS framework. The regularizer is build upon the group lasso with nested groups (see e.g., \citealp{nicholson2020high}) and encourages a hierarchical sparsity pattern that prioritizes the inclusion of coefficients according to the recency of the information the corresponding series contains about the state of the economy. 

We consider two empirical applications. First, we consider HF volatility forecasting through LF macroeconomic variables. The relationship between financial volatility and macroeconomic or financial variables has been analyzed in the literature, including the mixed-frequency context (e.g., \citealp{engle2013stock}; \citealp{asgharian2013importance}; \citealp{conrad2015anticipating}; \citealp{conrad2020two}; \citealp{fang2020predicting}). Most studies find that macroeconomic variables only pay off  when it comes to long-term forecasting but not so much for short-term horizons. We contribute to this literature by assessing the performance of the proposed regularized pooled RU-MIDAS model in forecasting the daily realized volatility of the S\&P 500 through a diverse collection of monthly macroeconomic variables. 
We find that the pooled RU-MIDAS model leads to (slightly) better forecast performance compared to the original RU-MIDAS model. However, the addition of macroeconomic variables does not improve forecast accuracy of the daily S\&P 500 realized variance compared to the popular HAR model of \cite{corsi2009simple}. 

In our second empirical application, we
consider HF bicycle rental demand forecasting through LF ridership variables. To this end, we
use publicly available  hourly demand rental data from a bicycle-sharing system in New York City for the year 2023. We forecast the hourly demand for bicycles using publicly available daily ridership data on traffic estimates from other transportation types in New York City. To the best of our knowledge, such mobility application has not yet been explored in the mixed-frequency literature. We find that the pooled RU-MIDAS model leads to better forecast performance compared to natural univariate benchmarks that only use past information of the HF target variable. These forecast gains are noticeable for short intra-day forecast horizons, and then gradually disappear when considering day-ahead forecasting.

The remainder of this paper is structured as follows. Section \ref{section:RU-MIDAS model}  introduces the pooled RU-MIDAS with hierarchically structured parameters and describes the regularized estimation procedure. Section \ref{section:empirical application:volatility} presents the results on the empirical application for volatility forecasting, Section \ref{section:empirical application:ridership} on the empirical application for bicycle rental forecasting. 
Section \ref{section:conclusion} concludes. Additional results are available in the Appendix.

\section{Regularized Estimation of the Pooled RU-MIDAS Model}\label{section:RU-MIDAS model}

We start by revising the approximate RU-MIDAS  regression of \cite{foroni2018rmidas} based on linear lag polynomials in Section \ref{subsec:RUMIDAS}.\footnote{We slightly deviate from \citeauthor{foroni2018rmidas}'s (\citeyear{foroni2018rmidas}) notation to facilitate the introduction of hierarchical structures in Section \ref{section:Hierarchical_Structures}. First, we make the LF variable observable at the end of LF period, together with the last HF observation within the LF period. In contrast, in \cite{foroni2018rmidas} the LF variable is observable at the beginning of the LF period together with the first HF observation within the LF period. Second, we define the dummy variables in the dummy notation (see equation \eqref{eq:RUMIDAS_dummy}) differently. Third, the first lag of the LF variable always corresponds to the previous LF period.} Then, we introduce the pooled RU-MIDAS model in Section \ref{subsec:pooling}, followed by  the hierarchical sparsity structure on its parameters in Section \ref{section:Hierarchical_Structures}. Finally, the regularized estimator for the pooled RU-MIDAS model is presented in Section \ref{subsec:estimator}.

\subsection{RU-MIDAS Model} \label{subsec:RUMIDAS}
\begin{figure}[t]
    \begin{subfigure}[t]{0.45\textwidth}
    \resizebox{\linewidth}{!}{
   \begin{tikzpicture}[
    roundnode/.style={circle, draw=black, fill=white, very thick, minimum size=7mm},
    squarednode/.style={rectangle, draw=black, fill=white, very thick, minimum size=11mm},
    ]
    \draw (0,0) -- (9,0);
        \foreach \x in {0,3,6,9}
          \draw (\x cm,3pt) -- (\x cm,-3pt);
    \draw (0,0) node[below=5pt](Q) {Q$_{t-1}$};
    \draw (3,0) node[below=5pt](M1) {M1$_{t}$};
    \draw (6,0) node[below=5pt](M2) {M2$_{t}$};
    \draw (9,0) node[below=5pt](M3) {M3$_{t}$};
    
    \draw[-, color={black}, draw opacity=1, line width=0.5mm] (0,0.2) to [out=40,in=140] (3,0.2);
    \draw[-, color={black}, draw opacity=1, line width=0.5mm] (3,0.2) to [out=40,in=140] (6,0.2);
    \draw[->, color={black}, draw opacity=1, line width=0.5mm] (6,0.2) to [out=40,in=140] node[above = 0.2pt] {$\alpha_3$} (9,0.2);
    \draw[-, color={black}, draw opacity=1, line width=0.5mm] (Q.south) to [out=320,in=220] (M1.south);
    \draw[->, color={black}, draw opacity=1, line width=0.5mm] (M1.south) to [out=320,in=220] node[below = 0.2pt] {$\alpha_2$} (M2.south);
    \draw[->, color={black}, draw opacity=1, line width=0.5mm] (Q.south) to [out=330,in=215] node[above = 0.2pt] {$\alpha_1$} (M1);
    \end{tikzpicture}
    }
    \caption{Lagged quarterly effect}
    \end{subfigure} 
    \begin{subfigure}[t]{0.55\textwidth}
    \resizebox{\linewidth}{!}{
   \begin{tikzpicture}[
    roundnode/.style={circle, draw=black, fill=white, very thick, minimum size=7mm},
    squarednode/.style={rectangle, draw=black, fill=white, very thick, minimum size=11mm},
    ]
    \draw (0,0) -- (10,0);
        \foreach \x in {0,2,4,6,8,10}
          \draw (\x cm,3pt) -- (\x cm,-3pt);
    \draw (0,0) node[below=5pt](M1) {M1$_{t-1}$};
    \draw (2,0) node[below=5pt](M2) {M2$_{t-1}$};
    \draw (4,0) node[below=5pt](M3) {M3$_{t-1}$};
    \draw (6,0) node[below=5pt](M1t) {M1$_{t}$};
    \draw (8,0) node[below=5pt](M2t) {M2$_{t}$};
    \draw (10,0) node[below=5pt](M3t) {M3$_{t}$};

      \draw[->, color={black}, draw opacity=1, line width=0.45mm] (4,0.2) to [out=40,in=140] node[below = 0.1pt] {$\textcolor{orange}{\beta_{1,1}}$} (6,0.2);
     \draw[-, color={black}, draw opacity=1, line width=0.45mm] (2,0.4) to [out=40,in=140] node[below = 0.1pt] {$\textcolor{Turquoise}{\beta_{2,1}}$} (4,0.4);
    \draw[->, color={black}, draw opacity=1, line width=0.45mm] (4,0.4) to [out=40,in=140] (6,0.4);
    \draw[-, color={black}, draw opacity=1, line width=0.45mm] (0,0.6) to [out=40,in=140] node[below = 0.1pt] {$\textcolor{ForestGreen}{\beta_{3,1}}$} (2,0.6);
    \draw[-, color={black}, draw opacity=1, line width=0.45mm] (2,0.6) to [out=40,in=140]  (4,0.6);
    \draw[->, color={black}, draw opacity=1, line width=0.45mm] (4,0.6) to [out=40,in=140] (6,0.6);
      
    \draw[->, color={black}, draw opacity=1, line width=0.45mm] (6,-0.7) to [out=320,in=220] node[above = 0.1pt]{$\textcolor{orange}{\beta_{1,2}}$} (8,-0.7);
    \draw[-, color={black}, draw opacity=1, line width=0.45mm] (4,-0.9) to [out=320,in=220]node[above = 0.1pt]{$\textcolor{Turquoise}{\beta_{2,2}}$} (6,-0.9);
    \draw[->, color={black}, draw opacity=1, line width=0.45mm] (6,-0.9) to [out=320,in=220] (8,-0.9);
    \draw[-, color={black}, draw opacity=1, line width=0.45mm] (2, -1.1) to [out=320,in=220]node[above = 0.1pt]{$\textcolor{ForestGreen}{\beta_{3,2}}$}(4,-1.1);
    \draw[-, color={black}, draw opacity=1, line width=0.45mm] (4,-1.1) to [out=320,in=220] (6,-1.1);
    \draw[->, color={black}, draw opacity=1, line width=0.45mm] (6,-1.1) to [out=320,in=220] (8,-1.1);

    \draw[-,  color={black}, draw opacity=1, line width=0.45mm] (6, 0.8) to [out=40,in=140] node[below = 0.1pt] {$\textcolor{Turquoise}{\beta_{2,3}}$} (8, 0.8);
    \draw[->,  color={black}, draw opacity=1, line width=0.45mm] (8,0.8) to [out=40,in=140] (10,0.8);
    \draw[->, color={black}, draw opacity=1, line width=0.45mm] (8,0.6) to [out=40,in=140] node[below = 0.1pt] {$\textcolor{orange}{\beta_{1,3}}$} (10,0.6);
    \draw[-, color={black}, draw opacity=1, line width=0.45mm] (4,1) to [out=40,in=140]node[above = 0.1pt] {$\textcolor{ForestGreen}{\beta_{3,3}}$} (6,1);
    \draw[-, color={black}, draw opacity=1, line width=0.45mm] (6,1) to [out=40,in=140] (8,1);
    \draw[->, color={black}, draw opacity=1, line width=0.45mm] (8,1) to [out=40,in=140] (10,1);
    \end{tikzpicture}
    }
     \caption{Lagged monthly effect}
    \end{subfigure} 
\caption{Lagged dynamics of the RU-MIDAS model for the monthly/quarterly setting.} 
    \label{fig:time steps groups}
\end{figure}
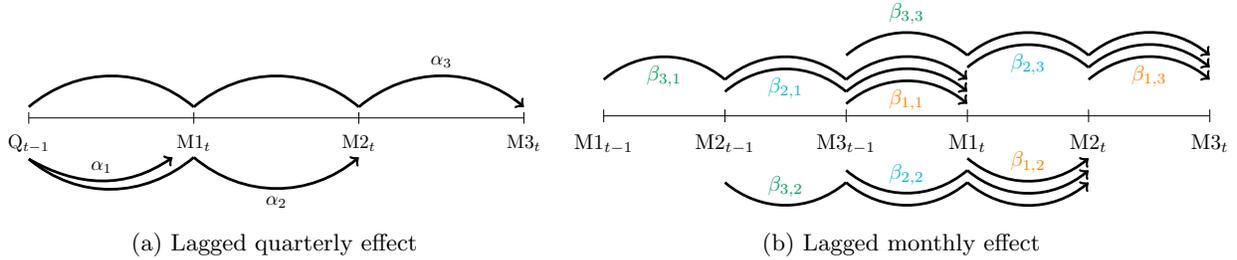

Let $x_t$ denote the high-frequency (HF) variable which we observe at $t = \frac{1}{m}, \frac{2}{m}, \dots, \frac{m-1}{m}, 1, 1 + \frac{1}{m}, ... $. and $y_t$ the low-frequency (LF) variable which is only observed every $m$ periods at $t=1,2,\dots, T$, where $m$ denotes the frequency mismatch. For simplicity, we consider one lag for the LF variable and $m$ lags for the HF variable.\footnote{We can easily extend the methodology to include multiple high and low frequency variables with higher-order lags at the cost of more complicated notation.} The  reverse unrestricted MIDAS (RU-MIDAS) model is given by
\begin{align}\label{eq:RUMIDAS_eqbyeq}
    x_{t} &= \alpha_{i} \ y_{t-\frac{i}{m}} + \beta_{1,i} \ x_{t-\frac{1}{m}} + \beta_{2,i} \ x_{t-\frac{2}{m}} + \dots + \beta_{m,i} \
    x_{t-1} + \varepsilon_{t}, \\
    t &= \frac{i}{m}, 1 + \frac{i}{m},\dots,
    \tau + \frac{i}{m}, \dots, 
    T - \frac{m-i}{m} \nonumber \\
    i &= 1,\dots,m, \nonumber
\end{align}
where $\alpha_i$, $\beta_{j,i}$ ($i,j = 1,\ldots,m$) are the unknown parameters, $\varepsilon_{t}$ the error term and the variables are assumed to be mean-centered such that no intercept is included. For illustrative purposes, consider equation \eqref{eq:RUMIDAS_eqbyeq} for the monthly/quarterly setting with $m=3$. Then, $\alpha_i$ and $\beta_{j,i}$ ($j = 1,2,3$) correspond to the effects of, respectively, the lagged quarterly and monthly variables on the $i$th ($i = 1,2,3$) monthly HF variable. Hence, the subscript $i$ of the coefficients refers to the effect of the corresponding regressor on month $i$, whereas $j$ indicates the number of months the monthly regressor is lagged by. For $i=1$ (first month so for $t = \frac{1}{3}, 1+\frac{1}{3},...$), $\alpha_1$ contains the effect of the previous quarter onto the first month and $\beta_{1,1},\beta_{2,1}$ and $\beta_{3,1}$ the effect of respectively month three, two and one of the previous quarter on the first month. Similarly for $i=2$ (second month so for $t = \frac{2}{3}, 1+\frac{2}{3},...$), $\alpha_2$ contains the effect of the previous quarter, however $\beta_{1,2}$ now contains the effect of month one of the current quarter, i.e., the latest available monthly observation (first monthly lag), and $\beta_{2,2},\beta_{3,2}$ refer to respectively month three and two of the previous quarter. For $i=3$ (third month so for $t = 1, 2,...$), the latest available monthly observations are month two ($\beta_{1,3}$) and one ($\beta_{2,3}$) of the current quarter, followed by month three of the previous quarter ($\beta_{3,3}$). The RU-MIDAS model thus easily allows for the inclusion of HF information released within the current LF period. Figure \ref{fig:time steps groups} provides a visual summary of these lagged dynamics for the RU-MIDAS model with the monthly/quarterly set-up. Finally, note that for ease of exposition, we include the same LF lag for each HF response in equation \eqref{eq:RUMIDAS_eqbyeq}, yet the model can easily accommodate different LF lags across HF responses to take, for instance, data releases into account. 

\cite{foroni2018rmidas} propose to group the system of equations in equation \eqref{eq:RUMIDAS_eqbyeq} into a single equation 
\begin{align} \label{eq:RUMIDAS_dummy}
    x_t &= \alpha_1 D_1 y_{t-\frac{1}{m}} + \alpha_2 D_2 y_{t-\frac{2}{m}} + \dots + \alpha_{m} D_{m} y_{t-1} \\
    &+ \beta_{1,1} D_1 x_{t-\frac{1}{m}} + \beta_{1,2} D_2 x_{t-\frac{1}{m}} +\dots+ \beta_{1,m} D_{m} x_{t-\frac{1}{m}} \nonumber \\
    &+ \beta_{2,1} D_1
    x_{t-\frac{2}{m}} + \beta_{2,2} D_2 x_{t-\frac{2}{m}} +\dots+ \beta_{2,m} D_{m} x_{t-\frac{2}{m}} \nonumber \\
    & \ \vdots  \nonumber \\
    &+ \beta_{m,1} D_1
    x_{t-1} + \beta_{m,2} D_2 x_{t-1}+\dots+ \beta_{m,m} D_{m} x_{t-1} + v_t \nonumber \\
    t&= \frac{1}{m},\frac{2}{m}, \dots, 1,\dots, T, \nonumber 
\end{align}
through the use of dummy variables $D_1$, $D_2$,\dots, $D_m$ taking respectively the value of one in the first ($t = \frac{1}{m}, 1+\frac{1}{m},...$), second ($t = \frac{2}{m}, 1+\frac{2}{m},...$), ... and $m$-th ($t = 1, 2,...$) HF observation within the LF period. We return to this dummy notation in Section \ref{subsec:pooling} due to its convenience for introducing the pooled RU-MIDAS model.

In a low-dimensional setting, the parameters of the RU-MIDAS model can be estimated by simple OLS. However, if the frequency mismatch $m$ is large, the model suffers from the curse of dimensionality since for one LF and $m$ HF lags $(1+m)m$ parameters need to be estimated with only $T$ low-frequency observations available, thereby making OLS unreliable. Indeed, first, the larger $m$ the more equations $i=1,\ldots,m$ need to be estimated which in turn increases the total number of parameters since each equation has its own set of parameters. Second, a large $m$ drastically decreases the sample size of the HF variable as the number of observations available for the estimation of each equation is limited to the number of LF observations $T$. The curse exacerbates if the number of LF and/or HF lags grows. Next, we propose to counter this sample size reduction by pooling corresponding lagged HF coefficients across equations.

\subsection{Pooled RU-MIDAS Model} \label{subsec:pooling}
To overcome the reduction in sample size, we propose to pool the coefficients $\beta_{j,i}$ of the lagged HF variable. For illustrative purposes, consider again equation \eqref{eq:RUMIDAS_eqbyeq} for the monthly/quarterly setting. Assume that the lagged effect of the previous month onto the current month is the same regardless of the HF period $i=1,2,3$. Then we can restrict the coefficients by pooling, thereby imposing the restrictions $\beta_{1,1} = \beta_{1,2} = \beta_{1,3}$. Similarly, we assume that the lagged two months effect onto the current month is the same regardless of the HF period, i.e., $\beta_{2,1} = \beta_{2,2} = \beta_{2,3}$ and correspondingly for the the third month $\beta_{3,1} = \beta_{3,2} = \beta_{3,3}$. In sum, re-consider Figure \ref{fig:time steps groups}(b) where the constraints imposed under pooling for the monthly/quarterly set-up are visualized through colour coding of the parameter labels, i.e., parameters with the same font color are pooled together. 

Generally, let us introduce the ``pooled RU-MIDAS" model in dummy notation as
\begin{align}\label{eq:RUMIDAS_pooled}
    x_{t} &= \alpha_1 D_1 y_{t-\frac{1}{m}} + \alpha_2 D_2 y_{t-\frac{2}{m}} + \dots + \alpha_{m} D_{m} y_{t-1} \\
    &+ \beta_{1} \ x_{t-\frac{1}{m}} + \beta_{2} \ x_{t-\frac{2}{m}} + \dots + \beta_{m} \
    x_{t-1} + v_{t}, \nonumber \\
    t &= \frac{1}{m}, \frac{2}{m}, \dots, 
    T. \nonumber 
\end{align}

Equation \eqref{eq:RUMIDAS_pooled} differs from \eqref{eq:RUMIDAS_dummy} in several aspects.
In the RU-MIDAS model \eqref{eq:RUMIDAS_dummy}, we consider a different equation for each HF variable which implies that the parameters capturing the lagged LF and HF effects vary for each HF component. In the pooled version \eqref{eq:RUMIDAS_pooled}, in contrast, we only allow the effect of the lagged LF variable to vary for each HF component, whereas the lagged HF effects are constrained to be the same. Therefore, we now have $Tm$ observations available for the estimation of the lagged HF coefficients $\beta_1,\dots,\beta_m$ instead of just $T$, the number of LF observations. 
Furthermore, we greatly reduce the total number of estimated parameters from $(1+m)m$ to $2m$. 
Since even the later number can be considerable if the frequency mismatch $m$ is large and/or not all $m$ lagged  HF or LF coefficients are necessarily of importance in modeling the HF variables, we further impose, in the next section, a hierarchical sparsity structure on the lagged parameters in equation \eqref{eq:RUMIDAS_pooled}.

\subsection{Hierarchical Sparsity in the Pooled RU-MIDAS}\label{section:Hierarchical_Structures}
To further reduce the dimensionality in equation \eqref{eq:RUMIDAS_pooled}, we resort to sparse estimation. While one could opt for ``patternless" sparsity through the lasso (\citealp{tibshirani1996regression}), we propose to use the dynamic structure of the pooled RU-MIDAS in equation \eqref{eq:RUMIDAS_pooled} as an additional source of information to guide the estimation. We use a hierarchical sparse estimator to encourage structured sparsity patterns appropriate to the context of RU-MIDAS regressions, similarly to the one introduced in \cite{hecq2022hierarchical} for stacked MF-VARs. 

Consider the hierarchical sparsity pattern that naturally arises in the autoregressive (AR) coefficients of a pooled RU-MIDAS regression. Let $\bm{\beta}$ be the vector collecting the HF coefficients in  equation \eqref{eq:RUMIDAS_pooled}, namely   $\bm{\beta} = [\beta_{1},..., \beta_{m}]^\prime$, and define $\bm{\alpha} = [\alpha_{1},..., \alpha_{m}]^\prime$ similarly for the LF coefficients. For each of the $2$ sub-vectors, we impose a hierarchical priority structure for parameter inclusion. We prioritize the inclusion of lagged coefficients according to the recency of the information the lagged time series contains relative to $x_t$, in line with the recency-based priority structure introduced in \cite{hecq2022hierarchical}. The more recent the information contained in the lagged HF or LF predictor, the more informative, and thus the higher its priority of inclusion. For instance, for the HF lags  in $\bm{\beta}$, the first lag $\beta_{1}$ contains the most recent information, followed by the second HF lag $\beta_{2}$, etc. Thus, for $j< j'$ we prioritize parameter $\beta_{j}$ over $\beta_{j'}$ in the model. Restrictions encouraging decay across lags are commonly used in the mixed-frequency literature, see for instance \cite{babii2022machine} for univariate or \cite{ghysels2016macroeconomics} for multivariate models. A similar recency structure arises in the coefficients of the LF variable. The first coefficient $\alpha_{1}$ contains the most recent information as it captures the effect of the previous LF period onto the first HF period which is followed by $\alpha_2$, the effect of the previous LF period onto the second HF period etc. Therefore, for $i< i'$ we prioritize parameter $\alpha_{i}$ over $\alpha_{i'}$. In the next section, we introduce the  regularized estimator that encourages such  hierarchical sparsity structures for the estimated parameters in the pooled RU-MIDAS model \eqref{eq:RUMIDAS_pooled}. Through our empirical studies in Sections \ref{section:empirical application:volatility} and \ref{section:empirical application:ridership}, we also demonstrate the flexibility of the regularized estimator in accommodating other hierarchical sparsity structures.

\subsection{Hierarchical Group Lasso for the Pooled RU-MIDAS} \label{subsec:estimator}
We use a nested group lasso (\citealp{zhao2009composite}) to attain the hierarchical structures detailed in Section \ref{section:Hierarchical_Structures}. Define $\bm{\beta}^{(j:m)} = [\beta_{j}, \dots, \beta_{m}]^\prime$ for $1 \leq j \leq m$. A nested structure then arises with $\bm{\beta}^{(1:m)} \supset \dots \supset \bm{\beta}^{(m:m)}$. Similarly, let $\bm{\alpha}^{(i:m)} = [\alpha_{i}, \dots, \alpha_{m}]^\prime$ for $1 \leq i \leq m$. We define the hierarchical group lasso estimator for the pooled RU-MIDAS model as

\begin{align}\label{eq:optimization problem}
    (\widehat{\bm{\alpha}},\widehat{\bm{\beta}})&= \underset{\bm{\alpha},\bm{\beta}}{\text{argmin}} 
   \Bigg\{\frac{1}{2}\sum_{t = \frac{1}{m}, \frac{2}{m},...}^{T}
    \Big ( x_{t} - \sum_{i=1}^{m} \alpha_{i} D_i y_{t-\frac{i}{m}} - \sum_{j=1}^{m}\beta_{j} x_{t-\frac{j}{m}}
    \Big )^2 + \lambda \cdot\mathcal{P}^{\text{Hier}}(\bm{\alpha},\bm{\beta}) \Bigg \},
\end{align}
where $\lambda \geq 0$ is a tuning parameter and $\mathcal{P}^{\text{Hier}}(\bm{\alpha},\bm{\beta})$ denotes the hierarchical group penalty given by 
\begin{align} \label{eq:penalty}
   \mathcal{P}^{\text{Hier}}(\bm{\alpha},\bm{\beta}) = \sum_{i=1}^{m} \lVert  \bm{\alpha}^{(i:m)}\rVert_2 + \sum_{j = 1}^{m} 
    \lVert  \bm{\beta}^{(j:m)}\rVert_2.
\end{align}
The hierarchical structure for the high-frequency lags
is built in through the condition that if $\bm{\beta}^{(j:m)} = \bf{0}$, then $\bm{\beta}^{(j':m)} = \bf{0}$ where $j < j'$, equivalently for $\bm{\alpha}$. We apply the proximal gradient algorithm to efficiently solve the optimization problem in equation \eqref{eq:optimization problem}, as detailed in Appendix A of \cite{hecq2022hierarchical}. For the selection of the tuning parameter $\lambda$, we rely on the Bayesian Information Criterion (BIC), but we also conduct an additional analysis to investigate the sensitivity of our results to the usage of time series cross-validation  instead of the  BIC (see Section \ref{subsec:sensitivity:volatility}). Finally, to reduce the bias introduced by lasso-type estimators, we use a ``post-lasso" procedure where we apply the least squares on the variables retained by the hierarchical group lasso. This leads to improved empirical performance on our considered applications as discussed in Section \ref{subsec:sensitivity:volatility}.
We illustrate the performance of the pooled RU-MIDAS on  two empirical studies as discussed in the next sections.

\section{Volatility Application}\label{section:empirical application:volatility}

We consider the regularized estimator of the pooled RU-MIDAS to forecast the realized variance of the S\&P 500 using several LF macroeconomic variables. 
We first review the literature on volatility forecasting with macroeconomic variables in Section \ref{subsec:literature:volatility}.
We then discuss the data and our forecast set-up in Section \ref{subsec:data and forecast}.
Results are presented in Sections \ref{subsec:rumidas vs pooled rumidas} to \ref{subsec:sensitivity:volatility}.

\subsection{Literature Background}
\label{subsec:literature:volatility}
In finance, movements in volatility are closely tracked since they have a considerable impact on capital investment, consumption, and economic activities (see e.g., \citealp{tsay2005analysis}; \citealp{bauwens2012handbook}). 
Therefore, forecasting volatility and the understanding of its drivers is of interest to both academics and practitioners. In recent years, the relationship between financial volatility and macroeconomic variables has received increased attention in the literature, we contribute to this strand by investigating the predictive information contained in LF macroeconomic variables for forecasting HF financial volatility.

Two problems, however, that affect the construction of such volatility forecasts is that volatility is unobservable and that macroeconomic variables are usually only available at a lower frequency. To overcome these difficulties, \cite{engle2013stock} propose a GARCH-MIDAS model to bridge the gap between daily stock returns and low-frequency (e.g., monthly, quarterly) explanatory variables. This model has been extensively used in the literature (e.g., \citealp{asgharian2013importance}; \citealp{conrad2015anticipating}; \citealp{conrad2020two}; \citealp{fang2020predicting}), but the evidence on which macroeconomic variables have predictive ability for stock market volatility is mixed. Nevertheless, coincident/lagging indicators such as inflation and industrial production (\citealp{engle2013stock}) and the expectations about them (\citealp{conrad2015anticipating}), as well as leading indicators such as term spread and housing starts (\citealp{conrad2015anticipating}, \citealp{fang2020predicting}) seem to often improve forecast performance.
Additionally, the findings highlight that including macroeconomic variables pays off particularly when it comes to long-term forecasting and not so much for short-term horizons. 

While GARCH models treat volatility as latent, accurate nonparametric estimates of volatility have become available since the emergence of high-frequency data (see e.g., \citealp{andersen2001distribution}; \citealp{barndorff2002econometric}). Realized variances, computed as the sum of the squared intraday returns for a particular day, make volatility ``observable" and volatility forecast models relying on such data have become popular alternatives (see e.g., \citealp{hua2013forecasting}; \citealp{haugom2014forecasting}).\footnote{Apart from assessing forecast accuracy, one could explore Granger causality relations, see for instance \cite{gotz2016testing} who test Granger-causality from a HF variable, namely the SP500 bipower variation to a LF macro variable, namely industrial production.} In this context, the heterogeneous autoregressive (HAR) model (\citealp{corsi2009simple}) has become particularly popular because of its simplicity and good performance to capture long memory features and for forecasting. The HAR model explains the log-realized variance as a linear function of the log-realized variances of yesterday and the average over the last five days (last week) and last 20 days (last month) which we will refer to in the remainder as the \textit{day-week-month (dwm)} lag structure. 
The HAR model can  be seen as a  MIDAS regression model with step functions as a weighting scheme, since dimension reduction is achieved by imposing equality constraints on the coefficients in the autoregressive model of respectively lags 2 to 5 on the one hand and lags 6 to 20 on the other hand. The number of autoregressive parameters is thereby reduced from 20 (for the unrestricted AR(20)) to 3 (for the HAR).

Recently, it has become popular to compare the forecast performance of models with a more general lag structure selected by lasso-type estimators with the parsimonious dwm lag structure embedded in the HAR (e.g., \citealp{audrino2016lassoing}; \citealp{audrino2017testing}; \citealp{ding2021forecasting}; \citealp{wilms2021multivariate}). Though the results are mixed, there is some evidence that the dwm lag structure of the HAR model is hard to beat by more sophisticated lasso-based lag structures. So far, the addition of macroeconomic variables to such models has, however, not been explored yet. We build on this strand of the literature and investigate whether HF daily realized variances can be forecast more accurately by accounting for information in  LF macroeconomic variables. 
Our proposed model pooled RU-MIDAS model, in contrast to the HAR,  achieves dimension reduction by
(i) pooling lagged HF effects across HF equations, and 
(ii) encouraging hierarchical sparsity on both the LF and the HF lagged effects. We thus combine imposed equality restrictions with data-driven sparsity restrictions.

\subsection{Data Description and Forecast Set-up}\label{subsec:data and forecast}
\begin{figure}[t]
    \centering
    \includegraphics[width = \textwidth]{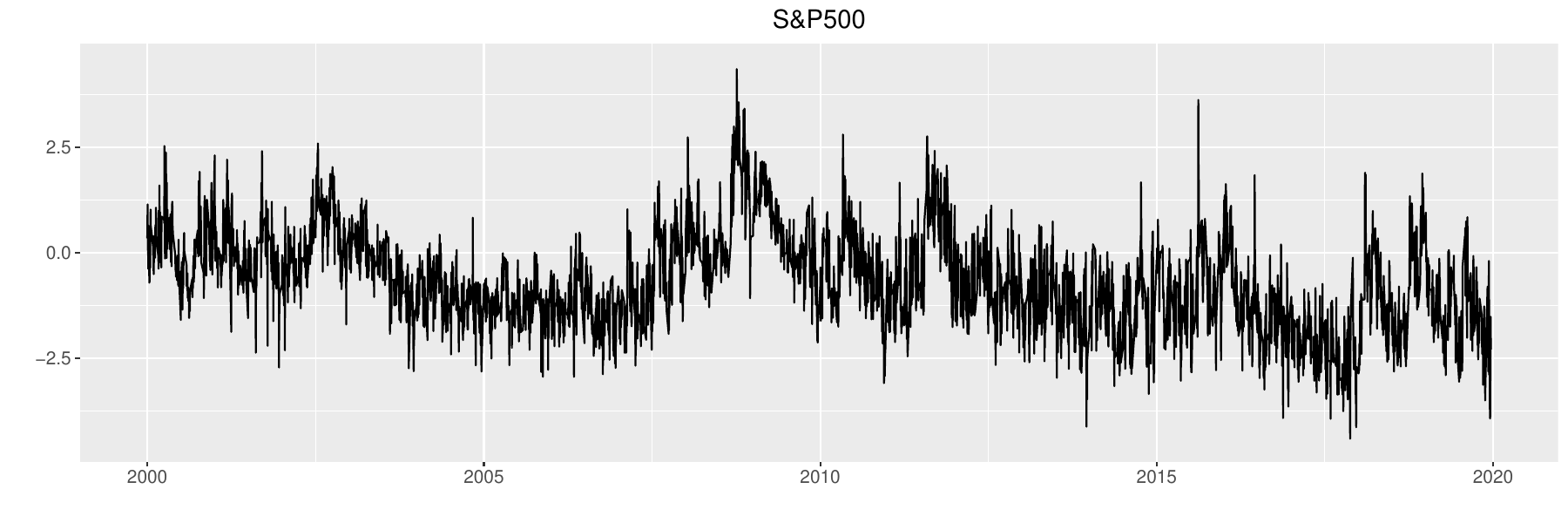}
    \caption{Log realized variances based on five-minute returns of the S\&P 500 from January 2000 to December 2019.} 
    \label{fig:SP500rv5}
\end{figure}

We forecast the daily realized variance of the S\&P 500 using a wide set of LF macroeconomic variables.
Regarding the HF variable, we consider the daily log realized variance based on five-minute returns of the Standard \& Poor’s 500 market index ``SP500" taken from the Oxford–Man Institute of Quantitative Finance.\footnote{Alternatively, we could have taken a more robust estimator of the integrated variance such as the bipower variation or the median realized volatility but this is not the focus of our paper.} The sample spans from January 3rd 2000 to December 31st 2019 to avoid the influence of the pandemic. Our mixed-frequency models require a fixed frequency mismatch $m$, which we set equal to $m=20$ as in \cite{gotz2016testing}. To ensure $m=20$ in each month, we therefore interpolate additional values for non-existing days for months with less than 20 trading days and disregard excessive days for months with more than 20 trading days at the beginning of the corresponding month. Figure \ref{fig:SP500rv5} plots the log-realized variances of the S\&P500. 

Regarding the LF macroeconomic variables, we consider monthly variables on various aspects of the economy: amongst others output, income, prices and employment, consumer sentiment and financial variables, see Table \ref{tab:data_description} for an overview. The monthly macroeconomic series are directly taken from the FRED-MD dataset which is available at the Federal Reserve Bank of St. Louis FRED database (see \citealp{mccracken2020real} for details). 
We apply the transformation codes in column ``T-code" to remove stochastic trends.\footnote{The transformation codes correspond to the codes in the FRED-MD dataset with the exception of ``CPIAUCSL" where we found the first log difference to be sufficient to remove the stochastic trend during the period we consider.} 
Rather than using all macroeconomic series in the FRED-MD dataset, we use a diverse yet careful selection of macroeconomic series with potential forecasting power for forecasting daily realized volatility.
Furthermore, we also consider the monthly news-based Economic Policy Uncertainty Index (``Epui") by \cite{baker2016measuring} as additional ``forward-looking" explanatory variable.
Finally, while some of the macroeconomic time series described in Table \ref{tab:data_description} are available at a higher frequency than monthly, we use monthly data for all these series to preserve the same daily-monthly mixed-frequency mismatch in our RU-MIDAS model.

\begin{table}[t]
    \centering
        \caption{Data description macroeconomic variables. Column T-code denotes the data transformation applied to a time series, which are: (1) not transformed, (2) $\Delta x_t$, (3) $ \Delta^2 x_t$, (4) $\text{log}(x_t)$, (5) $\Delta \text{log}(x_t)$, (6) $\Delta^2 \text{log}(x_t)$.}
        \resizebox{\textwidth}{!}{
                \centering
    \begin{tabular}{lllcccc}
    \hline
         Variable&FRED Code & Description &T-code  \\
        \hline
         Cpi&CPIAUCSL & CPI : All Items  & 5\\
         Csent & UMCSENTx & Consumer Sentiment Index (University of Michigan) & 2 \\
         Epui & - & News-based Economic Policy Uncertainty Index &1 \\
        Fedfunds&FEDFUNDS & Effective Federal Funds Rate & 2\\
        Houst & HOUST & Housing Starts: Total & 4 \\	
        Indpro & INDPRO & IP Index & 5 \\
         Oil&OILPRICEx & Crude Oil & 6\\
        M2& M2REAL & Real M2 Money Stock & 6\\	
        Termsp & GS10 - TB3MS &Term spread  (10-Year Treasury Rate (GS10) - 3-Month Treasury Bill (TB3MS)) & 2 \\
        Unrate&UNRATE & Civilian Unemployment Rate & 2\\
        Vix & VIXCLSx & VIX & 1 \\
         \hline
    \end{tabular}
    }
    \label{tab:data_description}
\end{table}

In our analysis, we consider 
eleven different pooled RU-MIDAS models 
where each model includes a different LF macroeconomic predictor.\footnote{We also assessed the predictive information from the joint model that includes all low-frequency variables, but this did not result in improved forecast performance.
Such a joint predictive model did, however, lead to a considerably higher computational burden.
Results are available from the authors upon request.} 
We include $20$ HF lags (one month) and one LF lag in the model.
To compare forecast performance of the different models and forecast methods, we perform a rolling window forecast exercise with a window size of 5 years ($T=1200$) and various forecast horizons ranging from daily ($h=1$), over weekly ($h=5$), and monthly ($h=20$), to several months ($h=40, 60, 120$).
To forecast more than one day-ahead in time (i.e., forecast horizon $h > 1$), we use a direct forecast approach (\citealp{marcellino2006forecast}). For a given forecast horizon $h$, the direct approach re-estimates model \eqref{eq:RUMIDAS_pooled} but with $x_{t+h-1}$ as the dependent variable. The advantage of using such direct approach is that it does not require the forecast of the lagged low-frequency variable. However, the model specification changes for each forecast horizon. Finally, forecast performance is measured based on the out-of-sample mean absolute forecast error (MAFE) but results with root mean squared forecast error (RMSFE) are available upon request. 

We have two main objectives in our empirical application that are subsequently investigated. First, we compare the forecast performance between the original RU-MIDAS and the proposed pooled RU-MIDAS model in Section \ref{subsec:rumidas vs pooled rumidas}, to investigate how the degree of parameter flexibility in the mixed-frequency model influences the forecast accuracy. Second, we analyze in Section \ref{subsec:pooled rumidas vs har} whether including LF macroeconomic variables for forecasting HF realized variance pays off by comparing the forecast performance between our pooled RU-MIDAS model and the HAR. We end with  sensitivity analyses in Section \ref{subsec:sensitivity:volatility}.

\subsection{Forecast Comparison RU-MIDAS and Pooled RU-MIDAS}\label{subsec:rumidas vs pooled rumidas}
In this section, we compare the forecast performance of the original RU-MIDAS and the proposed pooled RU-MIDAS model, where the parameters of both models are estimated using the hierarchical estimator for comparability. The estimation procedure for the pooled RU-MIDAS model is detailed in Sections \ref{section:Hierarchical_Structures} and \ref{subsec:estimator}, for the original RU-MIDAS model, we set hierarchical sparsity structures for each HF equation $i=1,\dots,m$ individually. For instance, for the high-frequency AR lags ($\bm{\beta}_i$) the first high-frequency  lag $\beta_{1,i}$ contains the most recent information, followed by the second HF lag $\beta_{2,i}$, etc. Thus, for $j< j'$ we prioritize parameter $\beta_{j,i}$ over $\beta_{j',i}$ in the model. As we only consider one LF lag and therefore $\alpha_i$ is a scalar, the group-lasso penalty for the LF lag boils down to a simple $\ell_1$-penalty. 

\begin{figure}[t]
    \centering
    \includegraphics[scale = 0.5]{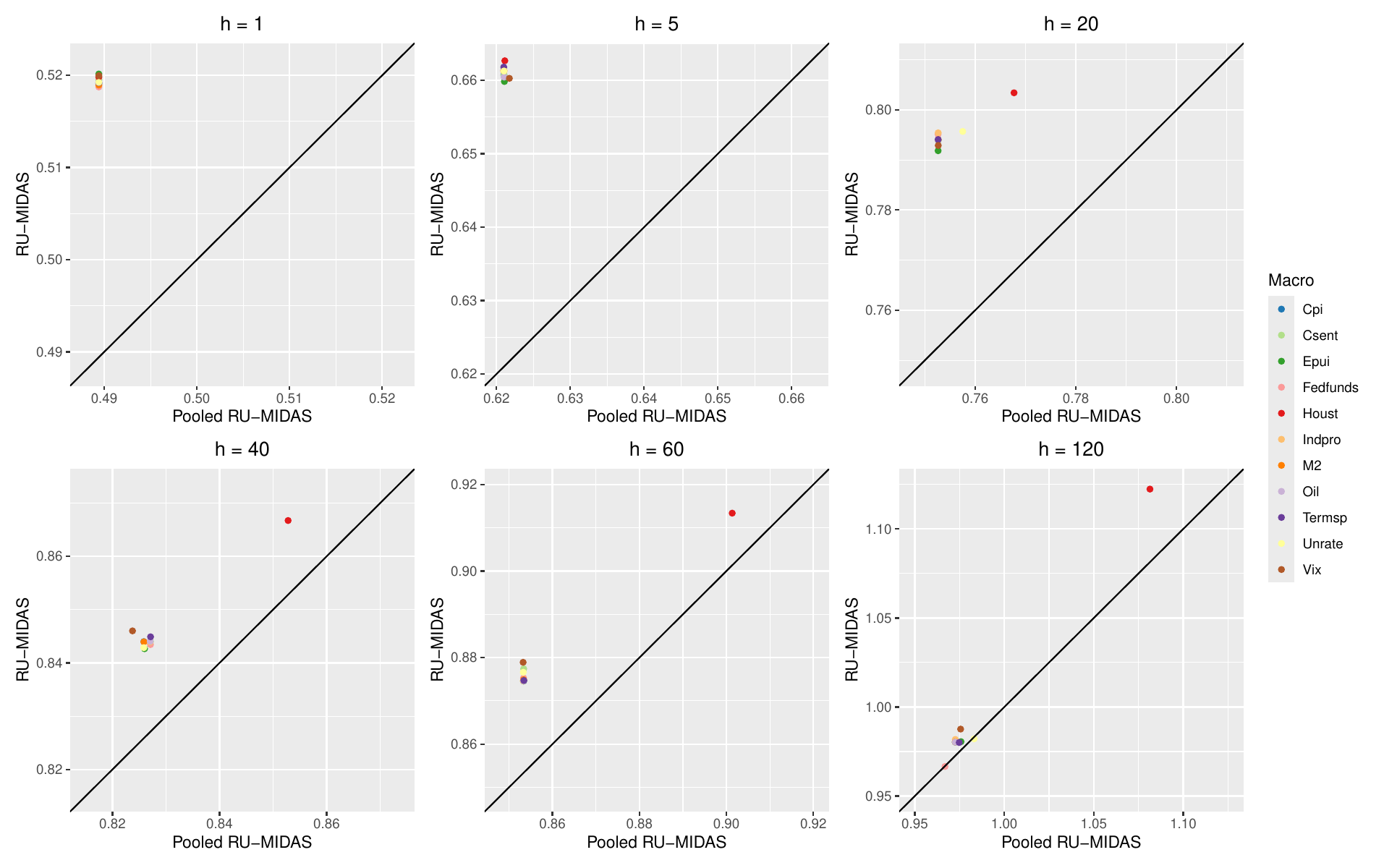}
    \caption{Volatility Application: Mean absolute forecast errors, averaged across time, of the pooled RU-MIDAS (x-axis) versus the original RU-MIDAS (y-axis) model, and this for the different macroeconomic variables (coloured dots) and the six forecast horizons (panels).}
    \label{fig:PooledRUM_vs_RUM}
\end{figure}

Figure \ref{fig:PooledRUM_vs_RUM} depicts the out-of-sample MAFEs for the pooled RU-MIDAS (x-axis) versus the original RU-MIDAS (y-axis) across the six forecast horizons (panels). Each panel includes 
eleven MAFEs (dots), one for each model with a particular macroeconomic variable as indicated in the coloured legend. The black line, being the 45 degree line, is added to facilitate comparison between the pooled RU-MIDAS model and the original one since dots above this line imply that the pooled model attains better average forecast accuracy (i.e. lower MAFEs). Across all forecast horizons (apart from $h=120$) and all macroeconomic variables, the pooled model delivers better forecast accuracy. This indicates that pooling corresponding lagged HF variables pays off for the considered data set. At short forecast horizons ($h=1, 5$),  the MAFEs are nearly indistinguishable across macroeconomic variables. 
For horizons $h=20, 40, 60, 120$,
the dot corresponding to ``Houst" (red) stands out. The model including Housing obtains higher MAFEs for both the pooled as well as the original RU-MIDAS model, hence the red dot is positioned in the upper right corner relative to the other dots. 

For each dot (i.e., each pairwise comparison between pooled RU-MIDAS and original RU-MIDAS) in Figure \ref{fig:PooledRUM_vs_RUM}, we perform a Diebold-Mariano (DM) test (\citealp{diebold1995comparing})  to test for significant difference in forecast accuracy. At horizon $h=1,5,20$, the pooled version is more accurate at 1\% significance level across all eleven 
models. At horizon $h=40, 60$, the pooled version performs better at (at least) 5\% significance level across all models with the exception of ``Houst" for which the improvement is statistically insignificant. 
For $h=120$, there is no significant improvement. In sum, we observe that the addition of the pooling constraints to the RU-MIDAS model   never leads to significantly worse forecast performance compared to the original RU-MIDAS model, often even better. We therefore continue with the pooled RU-MIDAS model and investigate whether the addition of  LF macroeconomic variables can improve forecast performance over the standard HAR model.

\subsection{Forecast Comparison Pooled RU-MIDAS and HAR}\label{subsec:pooled rumidas vs har}
In this section, we assess the predictive information  in the LF macroeconomic variables when forecasting the S\&P 500 daily realized variance. To this end, we compare the forecast performance between the pooled RU-MIDAS model, which includes macroeconomic variables, and the HAR that does not include macroeconomic variables. The parameters in the HAR model are estimated by OLS.

\begin{figure}[t]
    \centering
    \includegraphics[scale = 0.55]{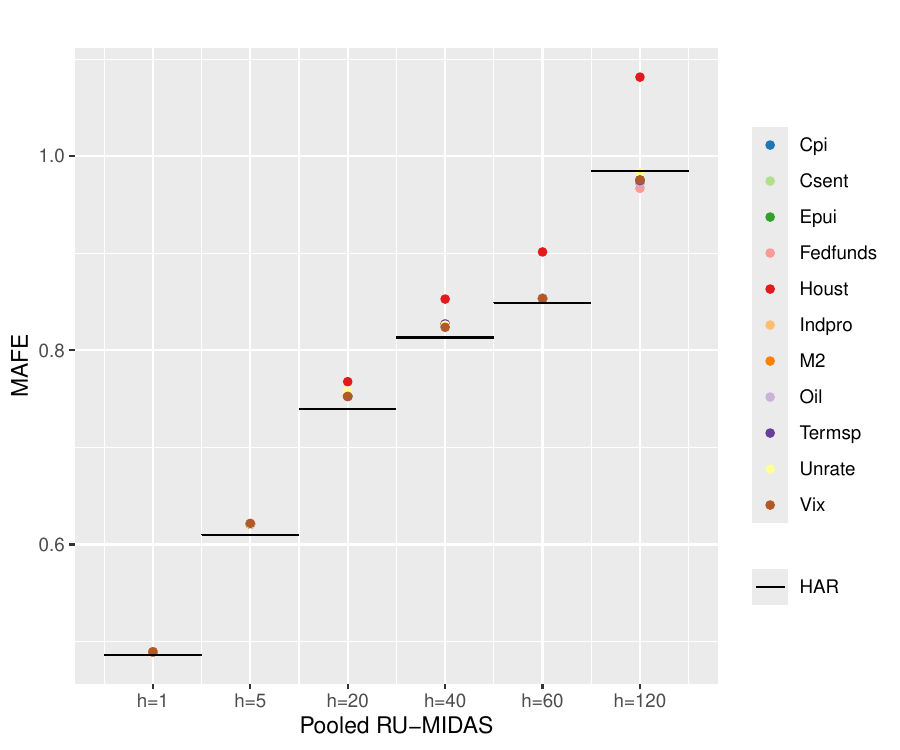}
    \caption{Volatility Application: Mean absolute forecast errors, averaged across time, of the pooled RU-MIDAS with different macroeconomic variables (dots) versus the HAR (solid black line) and this for the the six forecast horizons.
    }
    \label{fig:PooledRUM_vs_HAR}
\end{figure}

We first evaluate overall forecast performance, as computed via the MAFEs averaged across time. Figure \ref{fig:PooledRUM_vs_HAR} depicts the MAFE for the pooled RU-MIDAS estimated by the hierarchical estimator (dots, one for each macroeconomic variable) and the HAR (black line) across the six horizons. We clearly notice that, for both models, the forecast performance deteriorates with the forecast horizon (as expected). Moreover, the HAR consistently outperforms the pooled RU-MIDAS model by delivering lower MAFEs. Particularly, the addition of ``Houst" (red dot) deteriorates forecast accuracy in comparison to the other macroeconomic variables for which the difference is almost indistinguishable. 
Hence, overall, we find no evidence that the addition of macroeconomic variables improves forecast performance over the standard HAR. Similarly, \cite{fang2020predicting} observed that the inclusion of past volatility information  rather than the inclusion of macroeconomic variables improves forecasting accuracy. There are some exceptions at horizon $h=120$, namely for the models including the macroeconomic variables ``Epui", ``Fedfunds", ``Indpro", ``Termsp",``Unrate" and ``Vix"\footnote{The models with macroeconomic variables ``Cpi", ``Csent", ``M2", ``Oil" also obtain a lower MAFE than the HAR, however, no  macroeconomic lags are selected in these models (see Table \ref{tab:variable_selection} in the Appendix) so they only differ from the HAR in terms of their HF lag structure.}, which display a somewhat lower MAFE than the one obtained with the HAR model. It should, however, be noted that differences are largely negligible.

To further analyze the differences in forecast performance, Table \ref{tab:MAFE_MCS}  presents the MAFE of the HAR, and the pooled RU-MIDAS estimated by the hierarchical estimator (results as displayed in Figure \ref{fig:PooledRUM_vs_HAR}), as well as by OLS to better assess the influence of the regularization. For each forecast horizon, we consider these three estimation procedures, the two RU-MIDAS in combination with the 
eleven models, each including a different macroeconomic predictor, leading to 23 models in total. For each horizon, we obtain the models included in the 75\% Model Confidence Set (\citealp{hansen2011model}), whose MAFEs are displayed in bold.

\begin{table}[t]
\centering
\caption{Volatility Application: Mean absolute forecast errors, averaged across time, of the HAR, the pooled RU-MIDAS estimated by OLS (p-RU-MIDAS OLS) and the hierarchical estimator (p-RU-MIDAS HIER), and this for the different macroeconomic variables (columns) and the six forecast horizons. For each horizon, forecast methods in the 75\% Model Confidence Set (MCS) are in bold.}\label{tab:MAFE_MCS}
\makebox[\textwidth][c]{
\resizebox{1\textwidth}{!}{
\begin{tabular}{llccccccccccc}
  \hline
    Model &\multicolumn{11}{c}{Low-frequency variables} \\ \addlinespace[1pt]
& None & Cpi& Csent & Epui& Fedfunds & Houst& Indpro & M2 & Oil & Termsp & Unrate & Vix \\ 
\hline
 \rowcolor{lightgray} \multicolumn{13}{c}{Horizon $h=1$} \\
  \hline
  \addlinespace[2pt]
HAR & \textbf{0.4866}  \\ 
  p-RU-MIDAS OLS && \textbf{0.4904} & \textbf{0.4899} & \textbf{0.4909} & 0.4976 & \textbf{0.4907} & 0.4923 & 0.4930 & 0.4921 & 0.4927 & 0.4915 & 0.4927  \\ 
  p-RU-MIDAS HIER && \textbf{0.4894} & \textbf{0.4894}& \textbf{0.4894} & \textbf{0.4894} & \textbf{0.4894} & \textbf{0.4894} & \textbf{0.4894} & \textbf{0.4894} & \textbf{0.4894} & \textbf{0.4894} & \textbf{0.4894}  \\ 
  \addlinespace[2pt]
   \hline
  \rowcolor{lightgray} \multicolumn{13}{c}{Horizon $h=5$} \\
  \hline
  \addlinespace[2pt]
 HAR & \textbf{0.6097}\\ 
  p-RU-MIDAS OLS && 0.6170 & 0.6189 & 0.6170 & 0.6253 & 0.6166 & 0.6168 & 0.6208 & 0.6184 & 0.6183 & 0.6147 & 0.6191  \\ 
  p-RU-MIDAS HIER && 0.6210 & 0.6210 & 0.6211 & 0.6210 & 0.6211 & 0.6210 & 0.6210 & 0.6210 & 0.6210 & 0.6210 & 0.6217  \\ 
  \addlinespace[2pt]
   \hline
    \rowcolor{lightgray} \multicolumn{13}{c}{Horizon $h=20$} \\
  \hline
  \addlinespace[2pt]
 HAR & \textbf{0.7398} \\ 
  p-RU-MIDAS OLS && 0.7494 & \textbf{0.7506} & \textbf{0.7400} & \textbf{0.7565} & \textbf{0.7601} & \textbf{0.7473} & 0.7499 & \textbf{0.7455} & \textbf{0.7491} & \textbf{0.7433} & \textbf{0.7518}  \\ 
  p-RU-MIDAS HIER && \textbf{0.7526} & \textbf{0.7526} & \textbf{0.7526} & \textbf{0.7526} & \textbf{0.7677} & \textbf{0.7526} & \textbf{0.7526} &\textbf{ 0.7526 }& \textbf{0.7526 }& 0.7575 & \textbf{0.7527}  \\ \addlinespace[2pt]   \hline
 \rowcolor{lightgray} \multicolumn{13}{c}{Horizon $h=40$} \\
  \hline \addlinespace[2pt]
HAR & \textbf{0.8131} \\ 
  p-RU-MIDAS OLS && \textbf{0.8171} & \textbf{0.8229}& \textbf{0.8095} & \textbf{0.8201} & \textbf{0.8463} & \textbf{0.8188} &\textbf{ 0.8282} & \textbf{0.8184} & \textbf{0.8230} & \textbf{0.8208} & \textbf{0.8312} \\ 
  p-RU-MIDAS HIER && \textbf{0.8259} & \textbf{0.8271} & \textbf{0.8260} & \textbf{0.8271} & \textbf{0.8528} & \textbf{0.8270} & \textbf{0.8259} & \textbf{0.8271} & \textbf{0.8271} & \textbf{0.8259} & \textbf{0.8237} \\ \addlinespace[2pt]
   \hline
   \rowcolor{lightgray} \multicolumn{13}{c}{Horizon $h=60$} \\
\hline \addlinespace[2pt]
 HAR & \textbf{0.8487} \\ 
  p-RU-MIDAS OLS && \textbf{0.8535} & \textbf{0.8577} &\textbf{ 0.8448 }& \textbf{0.8438 }& \textbf{0.9007} &\textbf{ 0.8596} & \textbf{0.8630} & \textbf{0.8555} & \textbf{0.8512} & \textbf{0.8581} & \textbf{0.8850}  \\ 
  p-RU-MIDAS HIER && \textbf{0.8534} & \textbf{0.8534} & \textbf{0.8534} & \textbf{0.8534} & \textbf{0.9013} & \textbf{0.8533} & \textbf{0.8534} & \textbf{0.8534} & \textbf{0.8534} & \textbf{0.8533} & \textbf{0.8533} \\  
  \addlinespace[2pt]
  \hline 
  \rowcolor{lightgray} \multicolumn{13}{c}{Horizon $h=120$} \\
  \hline 
  \addlinespace[3pt]
  HAR & 0.9850 \\ 
  p-RU-MIDAS OLS && \textbf{0.9847 }& 0.9925 & \textbf{0.9728} & \textbf{0.9170} & \textbf{1.0880 }& 1.0185 &0.9997 & 0.9914 & \textbf{0.9547} & \textbf{1.0082} & 1.0829  \\ 
  p-RU-MIDAS HIER && \textbf{0.9726} & \textbf{0.9726} & \textbf{0.9758} & \textbf{0.9669} & \textbf{1.0814} & \textbf{0.9727} & \textbf{0.9726 }& \textbf{0.9726} & \textbf{0.9748 }&\textbf{ 0.9831} &\textbf{ 0.9756 } \\ \addlinespace[2pt]
  \hline 
\end{tabular}}}
\end{table}

We start by discussing the results for horizon $h=1$. Here, many combinations are included in the MCS with the HAR being the best performing one (on average). However, care is needed when evaluating the performance of the mixed-frequency models with macroeconomic variables included in the MCS vis-a-vis the HAR. Indeed,  Table \ref{tab:variable_selection} in the Appendix reveals that  the pooled RU-MIDAS models do not include any macroeconomic variables, since all their corresponding coefficients are set to zero by the regularized estimator. Hence, the HAR models and the pooled RU-MIDAS models only differ from each other in terms of the HF lag structure. The HAR model includes the HF lags via the parsimonious dwm lag structure, which cannot be improved upon for short forecast horizon. While the pooled RU-MIDAS model estimated by OLS includes all 20 HF lags,  the pooled RU-MIDAS model estimated by the hierarchical estimator includes on average around 4 HF lags (min. 3 lags, max. 6 lags) in the model. Our results confirm the previous findings by \cite{audrino2016lassoing},  \cite{audrino2017testing}, and \cite{wilms2021multivariate} on the limited evidence that the more general lag structure allowed by lasso-based approaches can beat the parsimonious day-week-month lag structure embedded in the HAR in terms of  out-of-sample forecast performance.

Across other forecast horizons, the HAR maintains its top performance. Hence, we do not find  evidence that the considered LF macroeconomic variables contain predictive information for forecasting the daily S\&P 500 daily realized variance. The only exception is $h=120$, where there is a small gain  in terms of average MAFE over the HAR when using the pooled RU-MIDAS models with various macroeconomic variables. It should, however, be noted that only in a minority of rolling windows (see Table \ref{tab:variable_selection} in the Appendix) this improved forecast accuracy over the HAR stems from the inclusion of macroeconomic variables.

Finally, let us discuss the performance of the pooled RU-MIDAS models estimated by OLS and the hierarchical estimator. We find that their performance is, overall, largely similar. The reason for that is that we employ a rolling window of five years to have sufficient low-frequency observations. However, with only one lag of a single low-frequency variable and twenty high-frequency lags included in the model, we need to estimate 40 coefficients while having a sample size of $T=1200$ high-frequency observations. Hence, the curse of dimensionality is rather limited in our set-up.

In sum, we find that macroeconomic variables do not help to improve the forecast performance of the daily S\&P 500 realized variance over the simple HAR. This finding is robust across a wide range of macroeconomic variables and forecast horizons. 
The superiority of the HAR model over more sophisticated statistical learning/machine learning methods is also confirmed by the recent study of \cite{audrino2024hard} including 1445 stocks from the US equity market. Indeed, the HAR model turns out to be exceptionally hard to beat especially  so when it is re-estimated daily in a rolling window set-up with window size around two and a half to four years.

\subsection{Sensitivity Analyses} \label{subsec:sensitivity:volatility}

{\it Rolling Window Size.} We assess the sensitivity of our findings to the choice of rolling window size. 
We re-did the forecast exercise using a rolling window size of 2 and 10 years instead of 5 years; the results are reported in respectively Table \ref{tab:MAFE_MCS_rw2} and \ref{tab:MAFE_MCS_rw10} in the Appendix. 
While the overall findings remain unchanged, we do observe, as expected, 
a 
more pronounced advantage of the regularized pooled RU-MIDAS model over the OLS-based one  if one were to use a smaller rolling window size of 2 years. 
For rolling window of 10 years, the advantage of using a regularized method in contrast to simple OLS largely disappears due to the sample size. Moreover, there is a minor benefit of including macroeconomic variables, as evidenced by the pooled RU-MIDAS estimated by OLS  often being included in the MCS together with the HAR.

{\it Tuning Parameter Selection.} We re-did the forecast exercise using time series cross-validation instead of the BIC to select the tuning parameter $\lambda$ in equation \eqref{eq:optimization problem}. We hereby use a rolling-window set-up, compute for each value of the tuning parameter the Mean Absolute Forecast Error (MAFE) across the last 20\% (i.e.\ one year) observations as cross-validation score, and select the value of the tuning parameter with the lowest MAFE. 
Results are reported in Tables \ref{tab:variable_selection_tscv} and \ref{tab:MAFE_MCS_sensitivity_tscv} of the Appendix.
Table \ref{tab:variable_selection_tscv}  reveals that the usage of time series cross-validation instead of BIC to select the tuning parameter leads to a considerably higher percentage of selected LF macroeconomic coefficients. Nonetheless, the addition of the LF macroeconomic variables in the model does not lead to improved out-of-sample forecast performance, as can be seen from Table \ref{tab:MAFE_MCS_sensitivity_tscv}. The MAFEs using either BIC or time series cross-validation are, overall, very similar.

{\it Post-lasso Estimation.} We re-did the forecast exercise without the post-lasso step to investigate its influence on the results. Table \ref{tab:MAFE_MCS_sensitivity_postlasso} in the Appendix reports the results for the pooled RU-MIDAS model both with and without post-lasso estimation.
We observe that the post-LASSO estimation step leads to improved forecast performance, and this across all considered models, macroeconomic variables and forecast horizons.

{\it High-frequency Lag Structure.} The pooled RU-MIDAS model is very flexible since it can easily incorporate other lag structures. 
For instance, one can easily incorporate the parsimonious dwm-lag structure used in the HAR into the penalized regression set-up of the pooled RU-MIDAS model. 
To this end, we re-did the forecast exercise using three different HAR-type models augmented with macroeconomic variables:
The 
(i) ``HAR+macro OLS"  procedure is  a HAR model (with dwm lag structure) augmented with a low-frequency macroeconomic variable in the pooled RU-MIDAS set-up and estimated through OLS. 
(ii) ``HAR+macro HIER" is the same model as above but estimation is done with the hierarchical regularizer. The hierarchical sparsity penalties can be applied with minimal adaptation to prioritize the daily lag over the weekly lag, which in turn could be prioritized over the monthly lag. We also use a sparsity penalty on the coefficients corresponding to the macroeconomic variables. 
(iii) ``HAR+macro HYBRID" is the same model as above but the HAR part of the regression is left unpenalized, and the  low-frequency macroeconomic predictors are penalized.
Table \ref{tab:MAFE_MCS_dwm} in the Appendix presents the MAFEs of these three procedures.
We do not find improved forecast performance over the standard HAR model, so our findings remain robust to the usage of the high-frequency lag structure in the pooled RU-MIDAS model.
Overall,  ``HAR+macro HYBRID" performs very similar to the HAR model since none of the low-frequency macroeconomic variables are selected (across most of the rolling windows). It outperforms ``HAR+macro HIER" (overall apart from $h=120$) which in turn  outperforms ``HAR+macro OLS".

\section{Ridership Application} \label{section:empirical application:ridership}
In our second empirical application, we use the pooled RU-MIDAS model to forecast demand for bicycle rentals in New York city using traffic estimates from several other transportation types.
While data from this bicycle-sharing system has been previously consider in \cite{hu2024fast,rombouts2024cross}, we are, to the best of our knowledge, the first to explore the predictive power of the ridership variables for  bicycle rentals through mixed-frequency models. In Section 
\ref{subsec:bike:data} we describe the data  together with our forecast set-up.
In Section \ref{subsec:bike:results}, we use the pooled RU-MIDAS model to forecast bicycle rental demand and present the results.

\subsection{Data Description and Forecast Set-up} \label{subsec:bike:data}
We forecast the hourly demand for bicycle rentals of the bicycle sharing system Citi Bike in New York City using a diverse set of LF ridership variables on various other transportation types.

\begin{figure}[t]
    \centering
    \includegraphics[width = 0.95\textwidth]{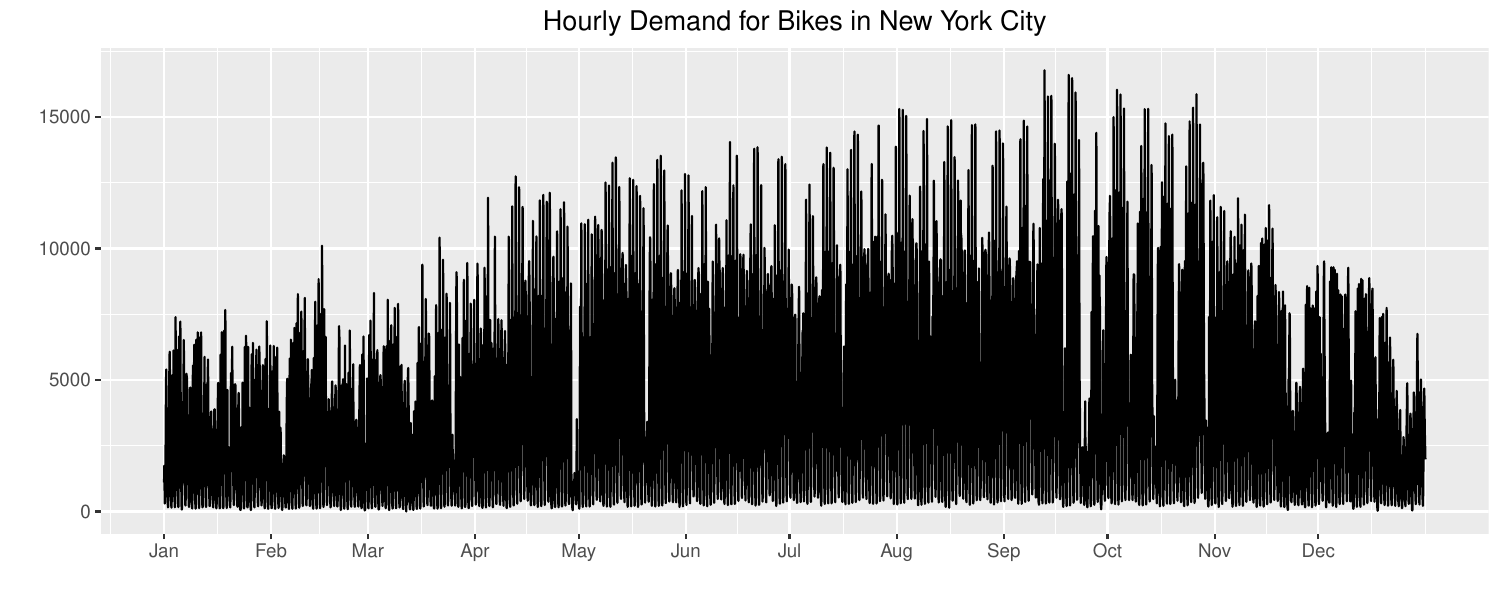}
    \includegraphics[width = 0.95\textwidth]{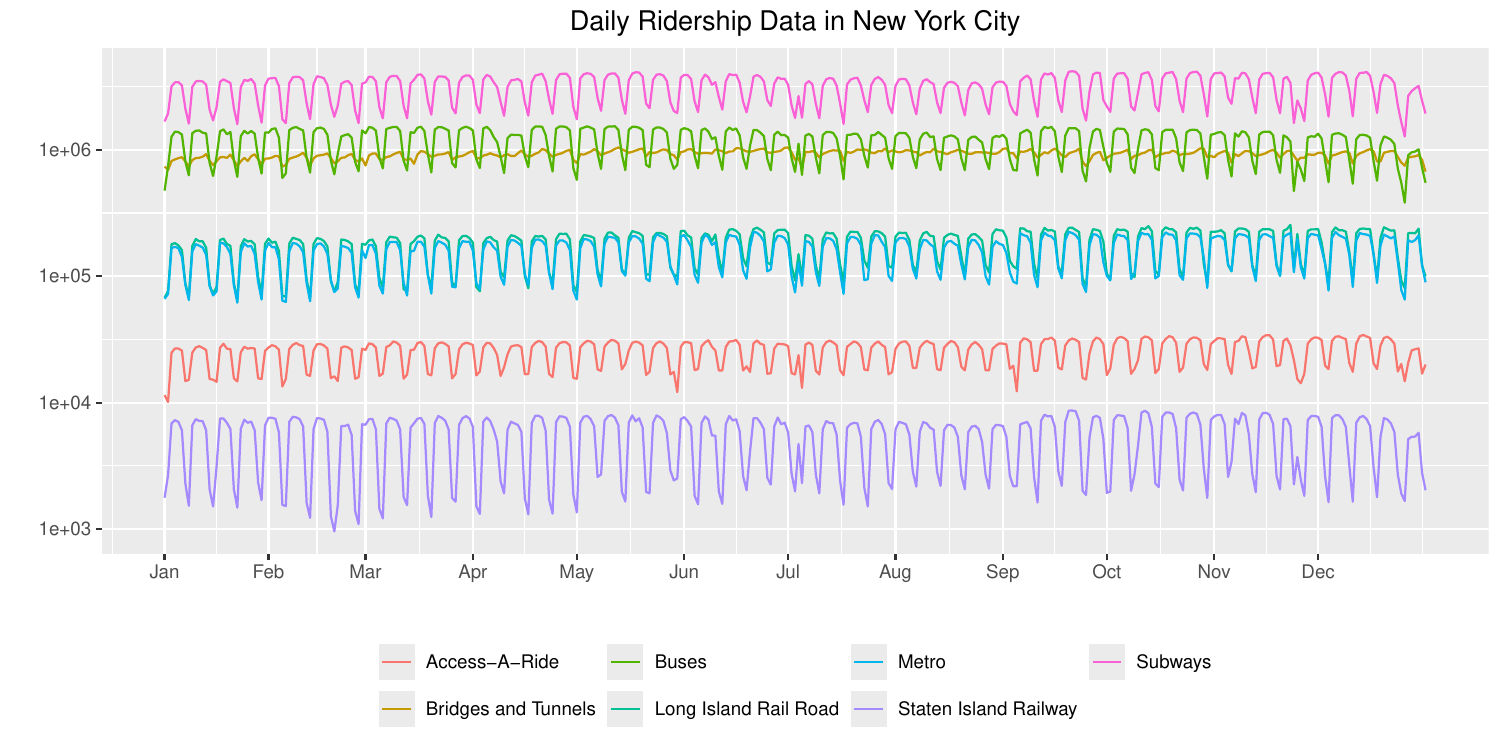}
    \caption{Hourly demand for bicycle rentals in New York City (Top) and daily traffic estimates for seven different transportation types (Bottom; on a log-scale) from January 2023 to December 2023.} 
    \label{fig:bikedemand}
\end{figure}

Regarding the HF variable, we collect publicly available hourly data from Citi Bike  which we process to obtain hourly time series measuring demand for bicycle rentals in New York City. The sample size spans January 1st 2023 to December 31st 2023, leading to a total of $T=8760$ hourly observations. 
Figure \ref{fig:bikedemand}, top panel, displays the hourly demand for bicycle rentals.
Although not directly visible from Figure \ref{fig:bikedemand}, there is clear intra-day as well as day-of-the-week seasonality present in the data.

Regarding the LF variables, we collect publicly available daily ridership data on traffic estimates for seven different transportation types in New York City provided by the Metropolitan Transportation Authority, see Table \ref{tab:data_description_ridership} for an overview.
Figure \ref{fig:bikedemand}, bottom panel, displays the daily traffic totals over the same sample period and this for the seven different ridership types on a log-scale.
The day-of-the-week seasonality is apparent from Figure \ref{fig:bikedemand}. 

In our analysis, we consider seven different pooled RU-MIDAS models with mixed frequency mismatch $m=24$ where each model includes a different LF ridership predictor.
We include $24\times7 = 168$ HF lags (one week) and 7 LF lags (one week) in the model.
To accommodate the seasonal patterns present in the bicycle and ridership data, the recency-based hierarchy of the penalty terms in equation \eqref{eq:penalty} can be easily adjusted to a seasonal-based hierarchy structure. In particular, for the HF lags, we prioritize the inclusion of the lag corresponding to the same hour of the previous week before the inclusion of all other lags according to their recency.
Similarly, regarding the LF lags, we prioritize the inclusion of the last week lag before the inclusion of the other lags according to their recency.
We consider a similar rolling-window forecasting set-up as the one described in Section \ref{subsec:data and forecast}. We  use a rolling window size of 30 days and consider  the following forecast horizons: $h=1,2,3,6,12$ hour(s)-ahead to cover intra-day forecast horizons  and finally $h=24$ to cover day-ahead forecasting. 

\begin{table}[t]
    \centering
        \caption{ Data description ridership variables.}
                \centering             
    \begin{tabular}{lllllll}   \hline

         Variable&&& Description   \\
        \hline
         Subways &&& Daily total estimated subway ridership \\
         Buses &&& Daily total estimated bus ridership \\
         LIRR &&& Daily total estimated Long Island Rail Road ridership \\
         Metro &&& Daily total estimated Metro-North ridership\\
         Access-A-Ride &&& Daily total scheduled Access-A-Ride trips\\
         Bridges and Tunnels &&& Daily total Bridges and Tunnels traffic \\             
         Staten Island Railway &&& Daily total estimated Staten Island Railway ridership\\          
         \hline
    \end{tabular}
    \label{tab:data_description_ridership}
\end{table}

We compare the forecast performance of the pooled RU-MIDAS models to two natural univariate benchmarks tailored towards the seasonality in the bicycle rental data. 
First, we use a seasonal version of the random walk (RW) model where the forecast of a specific hourly time slot is the value observed in the same hour and day of the  previous week.
This is a popular forecasting model for platform applications since it works well in the case of strong seasonality patterns as observed in our data.
As a second univariate benchmark model, we consider SARIMA models as implemented in the \texttt{auto.arima} function of the \texttt{forecast} package \citep{Rforecast-ref1, Rforecast-ref2} in \texttt{R} \citep{Rcoreteam}.  
We search the optimal model for our hourly bike rental series in the space of seasonal ARIMA models with seasonal frequency $s=24$.

\subsection{Results Forecast Performance} \label{subsec:bike:results}
Table \ref{tab:MAFE_MCS_ridership} presents the MAFE of the seasonal RW, the SARIMA model, and the pooled RU-MIDAS model estimated by either OLS or the hierarchical estimator. For each horizon, we obtain the models included in the 75\% MCS whose MAFEs are displayed in bold.

We first discuss the results for $h=1$. Both versions of the pooled RU-MIDAS model  outperform the seasonal RW model and the SARIMA model, and this regardless of the ridership type. Hence, there is predictive information contained in the LF ridership data for the HF demand of bicycle rentals.
Table \ref{tab:variable_selection_ridership} also confirms that across all ridership types, the LF ridership variables are selected (i.e.\ their corresponding coefficients are estimated as non-zero) in a large majority of the rolling windows, namely more than 85\% apart from the Bridges and Tunnels ridership type where they are still included in more than 65\% of the rolling windows.
The hierarchical estimator leads to the best forecast performance across all ridership types, and the pooled RU-MIDAS model with the ridership types Metro and Staten Island Railway are the best performing ones based on the MCS. 

\begin{table}
\centering
\caption{Ridership Application: Mean absolute forecast errors, averaged across time, of the seasonal RW, SARIMA, and the pooled RU-MIDAS estimated by OLS (p-RU-MIDAS OLS) and the hierarchical estimator (p-RU-MIDAS HIER), and this for the different ridership variables (columns) and the six forecast horizons. For each horizon, forecast methods in the 75\% Model Confidence Set (MCS) are in bold. }\label{tab:MAFE_MCS_ridership}
\makebox[\textwidth][c]{
\begin{tabular}{lcccccccc}
  \hline
    Model &\multicolumn{8}{c}{Low-frequency variables} \\ 
& None & Subways & Buses & LIRR &  Metro & \makecell{Access-\\A-Ride} & \makecell{Bridges \\ And\\Tunnels} & \makecell{Staten \\Island \\Railway} \\ 
\hline
 \rowcolor{lightgray} \multicolumn{9}{c}{Horizon $h=1$} \\
 \hline \addlinespace[2pt]
    seasonal RW & 978.67 \\
    SARIMA  & 471.08 \\
    p-RU-MIDAS OLS && 456.58 & 452.52 & 449.97 & 454.23 & 454.69 & 507.14 & 450.55 \\ 
    p-RU-MIDAS HIER && 426.22 & 427.14 & 424.76 & \textbf{422.30} & 433.16 & 467.88 & \textbf{419.47} \\  \addlinespace[2pt]
   \hline
 \rowcolor{lightgray} \multicolumn{9}{c}{Horizon $h=2$} \\
 \hline  \addlinespace[2pt]
seasonal RW & 978.90  \\ 
  SARIMA  & 855.13  \\ 
  p-RU-MIDAS OLS && 760.79 & 753.45 & 745.66 & 756.40 & 759.35 & 832.63 & 751.37 \\ 
  p-RU-MIDAS HIER && 711.39 & 718.50 & \textbf{704.24} & \textbf{698.77} & 720.11 & 782.78 & \textbf{699.26} \\  \addlinespace[2pt]  
\hline
 \rowcolor{lightgray} \multicolumn{9}{c}{Horizon $h=3$} \\
 \hline  \addlinespace[2pt]
seasonal RW & 979.14  \\ 
  SARIMA  & 1087.08  \\ 
  p-RU-MIDAS OLS && 936.74 & 931.39 & \textbf{919.06} & 932.66 & 935.58 & 1015.79 & 927.58 \\ 
  p-RU-MIDAS HIER && \textbf{879.91} & \textbf{881.33} & \textbf{873.59} & \textbf{869.43} & \textbf{883.93} & 913.01 & \textbf{872.91} \\ 
  \addlinespace[2pt]  
\hline
\rowcolor{lightgray} \multicolumn{9}{c}{Horizon $h=6$} \\
\hline  \addlinespace[2pt]
seasonal RW & \textbf{979.08}  \\ 
  SARIMA  & 1300.88  \\ 
  p-RU-MIDAS OLS && 1157.90 & 1141.50 & 1131.75 & 1152.34 & 1159.30 & 1216.69 & 1141.43 \\ 
  p-RU-MIDAS HIER && \textbf{987.18} & \textbf{984.06} & \textbf{983.32} & \textbf{985.33} & 995.20 & \textbf{989.36} & \textbf{983.58} \\  \addlinespace[2pt]
\hline
\rowcolor{lightgray} \multicolumn{9}{c}{Horizon $h=12$} \\
\hline  \addlinespace[2pt]
seasonal RW & \textbf{980.17}  \\ 
  SARIMA  & 1337.53  \\ 
  p-RU-MIDAS OLS && 1218.79 & 1201.75 & 1205.23 & 1228.81 & 1227.43 & 1267.15 & 1196.78 \\ 
  p-RU-MIDAS HIER && \textbf{944.43} & \textbf{942.83} & \textbf{943.66} & \textbf{942.63} & \textbf{943.45} & \textbf{944.78} & \textbf{943.25} \\  \addlinespace[2pt]
\hline
\rowcolor{lightgray} \multicolumn{9}{c}{Horizon $h=24$} \\
\hline  \addlinespace[2pt]
seasonal RW & 980.88  \\ 
  SARIMA  & 1347.56  \\ 
  p-RU-MIDAS OLS && 1251.32 & 1248.29 & 1237.84 & 1242.47 & 1275.40 & 1304.81 & 1236.59 \\ 
  p-RU-MIDAS HIER && \textbf{926.98} & \textbf{926.98} & \textbf{926.98} & \textbf{926.98} & \textbf{926.98} & \textbf{926.98} & \textbf{926.98} \\  \addlinespace[2pt]
  \hline
\end{tabular}}
\end{table}

\begin{table}
\centering
\caption{Ridership Application: Percentage of rolling windows in which at least one low-frequency ridership coefficient was selected (i.e.\ estimated as nonzero) across the six forecast horizons (rows). }\label{tab:variable_selection_ridership}
\makebox[\textwidth][c]{
\begin{tabular}{lccccccc}
  \hline
 Horizon & Subways & Buses & LIRR &  Metro & \makecell{Access-\\A-Ride} & \makecell{Bridges \\ And\\Tunnels} & \makecell{Staten \\Island \\Railway} \\ \addlinespace[2pt] 
 \hline
 $h=1$ & 85.73 & 83.18 & 86.12 & 87.91 & 82.80 & 68.80 & 88.59 \\ 
 $h=2$ &  57.06 & 54.52 & 58.22 & 60.19 & 54.49 & 45.95 & 60.53 \\ 
 $h=3$ & 28.81 & 24.02 & 29.42 & 30.29 & 26.35 & 26.69 & 30.52 \\ 
 $h=6$ &  3.79 & 3.59 & 3.28 & 2.90 & 3.73 & 9.44 & 3.46 \\ 
 $h=12$ & 0.52 & 0.56 & 0.38 & 0.08 & 0.15 & 0.34 & 0.33 \\ 
 $h=24$ &  0.00 & 0.00 & 0.00 & 0.00 & 0.00 & 0.00 & 0.00 \\ 
\hline
\end{tabular}}
\end{table}

Turning to the other intra-day forecast horizons ($h=2,3,6,12$), we notice that the pooled RU-MIDAS model with hierarchical estimator still outperforms the univariate benchmarks, although the margin by which it outperform these benchmarks, especially so the seasonal RW, becomes smaller as the forecast horizon increases. 
Table \ref{tab:variable_selection_ridership} reveals that the predictive power of the LF ridership variables gradually disappears as the forecast horizons increases, since fewer and fewer ridership variables are included in the pooled RU-MIDAS model. The selection percentages drop from more than 50\% for horizon $h=2$ to less than 1\% for horizon $h=12$. The MCS results reveal that the pooled RU-MIDAS models with hierarchical estimator still delivers the best performing models for horizons $h=2$ (LIRR, Metro and Staten Island Railway) and $h=3$ (all ridership types except for Bridges and Tunnels; the OLS pooled RU-MIDAS model with LIRR is also included in the MCS). For horizons $h=6$ and $h=12$, in contrast, the best performing pooled RU-MIDAS models perform equally good as the seasonal RW model.

Finally, for the day-ahead forecast horizon $h=24$, the pooled RU-MIDAS models still maintain their excellent performance when inspecting their MAFEs, yet the MAFEs are identical across ridership types, and Table \ref{tab:variable_selection_ridership} further reveals that the good forecast performance stems from the sole inclusion of the HF lags selected in the model, since no LF lags are selected.

In sum, the pooled RU-MIDAS model with hierarchical estimator offers clear forecast gains at short intra-day horizons over standard univariate benchmarks, but the predictive power of the LF ridership variables dies out quickly across larger forecast horizons. For day-ahead forecast horizons (or further), there is no advantage of using a mixed-frequency model over a standard univariate model for the HF series of interest.

\color{black}

\section{Conclusion}\label{section:conclusion}

We consider mixed-frequency RU-MIDAS regressions with a large frequency mismatch $m$ and propose two extensions to the framework to tackle their curse of dimensionality. We (i) pool the regression coefficients regulating the lagged HF dynamics across the $m$ different HF equations to increase the sample size and (ii) we use a hierarchical regularizer to discriminate between recent  and obsolete information. The RU-MIDAS model with hierarchical sparsity structures is quite flexible and can easily combine various high-frequency and low-frequency series in the model as well as incorporate other lag priority structures as demonstrated on two empirical applications. 

We first use the proposed regularized pooled RU-MIDAS model to forecast the daily realized volatility of the S\&P 500 through the information contained in a diverse set of monthly macroeconomic variables. While we find that the pooled RU-MIDAS model improves forecast performance upon the original (non-pooled) one, we do not find evidence that the monthly macroeconomic variables contain important information to improve forecast performance upon the popular benchmark HAR model that only  exploits past volatility information. 
We also use the pooled RU-MIDAS model to forecast hourly demand for bicycle rentals in New York City through the information contained in a diverse set of daily ridership variables of other transportation types.
Here, we do find evidence that the daily ridership variables contain important predictive power to improve forecast performance upon natural univariate benchmarks. This gain in forecast performance is visible for intra-day forecast horizons but not for day-ahead forecast horizons. It would be interesting to further explore the potential of regularized RU-MIDAS models for other application areas.

\paragraph{Data availability statement.}
Regarding the volatility application, the FRED-MD dataset is publicly available at \url{https://research.stlouisfed.org/econ/mccracken/fred-databases/}. The Economic Policy Uncertainty Index is available at \url{https://www.policyuncertainty.com}. The realized variance was taken from the Oxford–Man Institute of Quantitative Finance, however the public access has been terminated.
Regarding the ridership application, the Citi Bike dataset is publicly available at \url{https://citibikenyc.com/system-data}. The daily ridership data with traffic estimates is publicly available at \url{https://data.ny.gov}. Computer code is available from \url{https://github.com/MarieTernes/hierarchical-RUMIDAS}.

\bibliographystyle{chicago}
\bibliography{references}

\clearpage
\begin{appendices}
\section{Volatility Application: Additional Results}\label{appendix:rw}

\begin{table}[ht]
\centering
\caption{Percentage of rolling windows in which at least one low-frequency macroeconomic coefficient was selected (i.e.\ estimated as nonzero) across the six forecast horizons (rows).  }\label{tab:variable_selection}
\makebox[\textwidth][c]{
\begin{tabular}{lccccccccccc}
  \hline
 Horizon &Cpi& Csent & Epui &Fedfunds & Houst& Indpro & M2 & Oil & Termsp & Unrate & Vix  \\ 
 \hline
 $h=1$ & 0 & 0 & 0 & 0 & 0 & 0 & 0 & 0 & 0 & 0 & 0 \\
 $h=5$ &  0 &0 &0 & 0 &0.17& 0 &0&0 &0& 0& 0 \\
 $h=20$ &  0 &  0 & 0 & 0 &5.08 &  0 &  0  & 0  & 0 & 1.69 &  0 \\
 $h=40$ & 0 & 0 & 0 & 0& 4.57 & 0.31 & 0 & 0.43 & 0.57 & 0 & 0 \\ 
 $h=60$ &  0 & 0 & 0 & 0& 5.78 & 0 & 0 & 0 & 0 & 0& 0   \\
 $h=120$ &  0 & 0 & 3.44 & 10.74 &22.74& 0.57 & 0 & 0 & 0.87 & 1.20 & 0.60  \\
\hline

\end{tabular}}
\end{table}

\begin{table}[ht]
\centering
\caption{Sensitivity analysis on the tuning parameter selection with time series cross-validation. Percentage of rolling windows in which at least one low-frequency macroeconomic coefficient was selected (i.e.\ estimated as nonzero) across the six forecast horizons (rows). }\label{tab:variable_selection_tscv}
\makebox[\textwidth][c]{
\resizebox{1\textwidth}{!}{
\begin{tabular}{lccccccccccc}
  \hline
 Horizon &Cpi& Csent & Epui &Fedfunds & Houst& Indpro & M2 & Oil & Termsp & Unrate & Vix  \\ 
 \hline
 $h=1$ & 87.19 & 85.71 & 93.10 & 87.19 & 93.60 & 90.15 & 87.19 & 88.18 & 91.72 & 95.07 & 90.27  \\
 $h=5$ & 85.69 & 81.24 & 82.92 & 90.18 & 89.14 & 92.30 & 82.43 & 87.86 & 82.92 & 92.30 & 90.82 \\
 $h=20$ &  65.76 & 70.66 & 74.24 & 70.19 & 70.54 & 77.62 & 70.93 & 75.21 & 66.68 & 78.07 & 72.43 \\
 $h=40$ &  65.22 & 67.35 & 73.28 & 69.04 & 63.69 & 71.87 & 69.61 & 64.84 & 71.37 & 73.38 & 67.86\\ 
 $h=60$ &  74.63 & 69.56 & 78.69 & 77.68 & 70.57 & 72.27 & 68.54 & 74.63 & 73.69 & 80.90 & 70.07  \\
 $h=120$ &  71.17 & 68.08 & 72.79 & 89.40 & 75.69 & 76.98 & 73.84 & 81.63 & 70.04 & 75.07 & 70.85 \\
\hline

\end{tabular}}}
\end{table}

\begin{table}[h]
\centering
\caption{Sensitivity analysis for a rolling window of 2 years. Mean absolute forecast errors, averaged across time, of the HAR, the pooled RU-MIDAS estimated by OLS and the hierarchical estimator, and this for the different macroeconomic variables (columns) and the six forecast horizons. For each horizon, forecast methods in the 75\% Model Confidence Set (MCS) are in bold.}\label{tab:MAFE_MCS_rw2}
\makebox[\textwidth][c]{
\resizebox{1\textwidth}{!}{
\begin{tabular}{llccccccccccc}
  \hline
    Model &\multicolumn{12}{c}{Low-frequency variables} \\ \addlinespace[3pt]
& None & Cpi& Csent & Epui& Fedfunds & Houst& Indpro & M2 & Oil & Termsp & Unrate & Vix  \\ 
\hline 
 \rowcolor{lightgray} \multicolumn{13}{c}{Horizon $h=1$} \\
  \hline \addlinespace[3pt]
HAR & \textbf{0.4672} \\ 
  p-RU-MIDAS OLS && 0.4849 & 0.4833 & 0.4863 & 0.4890 & 0.4850 & 0.4851 & 0.4854 & 0.4867 & 0.4840 & 0.4833 & 0.4861  \\ 
  p-RU-MIDAS HIER && 0.4714 & 0.4714 & 0.4714 & 0.4714 & 0.4714 & 0.4714 & 0.4714 & 0.4714 & 0.4714 & 0.4714 & 0.4714  \\  \addlinespace[3pt]
  \hline
  \rowcolor{lightgray} \multicolumn{13}{c}{Horizon $h=5$} \\
  \hline \addlinespace[3pt]
HAR & \textbf{0.5812} \\ 
  p-RU-MIDAS OLS && 0.6071 & 0.6047 & 0.6054 & 0.6066 & 0.6019 & 0.6083 & 0.6036 & 0.6088 & 0.6041 & 0.6016 & 0.6042  \\ 
  p-RU-MIDAS HIER && 0.5973 & 0.5974 & 0.5973 & 0.5974 & 0.5973 & 0.5973 & 0.5973 & 0.5973 & 0.5973 & 0.5973 & 0.5973  \\  \addlinespace[3pt]
  \hline
 \rowcolor{lightgray} \multicolumn{13}{c}{Horizon $h=20$} \\
  \hline \addlinespace[3pt]
  HAR & \textbf{0.7154}\\ 
  p-RU-MIDAS OLS && 0.7395 & 0.7371 & 0.7333 & 0.7501 & 0.7532 & 0.7354 & 0.7308 & 0.7379 & 0.7405 & 0.7399 & 0.7440  \\ 
  p-RU-MIDAS HIER && 0.7264 & 0.7264 & 0.7264 & 0.7264 & 0.7264 & 0.7264 & 0.7264 & 0.7264 & 0.7264 & 0.7264 & 0.7264  \\ \addlinespace[3pt]
   \hline
   \rowcolor{lightgray} \multicolumn{13}{c}{Horizon $h=40$} \\
\hline \addlinespace[3pt]
HAR & \textbf{0.7940} \\ 
  p-RU-MIDAS OLS && \textbf{0.8111} & 0.8161 & 0.8236 & 0.8439 & \textbf{0.8305} & \textbf{0.8072} & \textbf{0.8186} & 0.8175 & \textbf{0.8170} & 0.8197 & \textbf{0.8204}  \\ 
  p-RU-MIDAS HIER && \textbf{0.7920} & \textbf{0.7920} & \textbf{0.7920} & \textbf{0.7920 }& \textbf{0.7920} & \textbf{0.7920} & \textbf{0.7920} & \textbf{0.7920} & \textbf{0.7920} & \textbf{0.7920} & \textbf{0.7920}  \\  \addlinespace[3pt]
   \hline
  \rowcolor{lightgray} \multicolumn{13}{c}{Horizon $h=60$} \\
  \hline \addlinespace[3pt]
  HAR & \textbf{0.8342}  \\ 
  p-RU-MIDAS OLS && 0.8807 & 0.8557 & 0.8737 & 0.8748 & \textbf{0.8722} & 0.8694 & 0.8791 & 0.8539 & 0.8639 & 0.8683 & 0.8878  \\ 
  p-RU-MIDAS HIER && \textbf{0.8321} & \textbf{0.8321} & \textbf{0.8321} & \textbf{0.8321} &\textbf{ 0.8273}& \textbf{0.8321} & \textbf{0.8321} & \textbf{0.8321} & \textbf{0.8321} & \textbf{0.8321} & \textbf{0.8321}  \\ \addlinespace[3pt]
  \hline
 \rowcolor{lightgray} \multicolumn{13}{c}{Horizon $h=120$} \\
 \hline \addlinespace[3pt]
  HAR & 1.0250 \\ 
  p-RU-MIDAS OLS && 1.0351 & 1.0312 & 1.0337 & \textbf{0.9833} & \textbf{1.0328} & 1.0889 & 1.0512 & 1.0417 & 1.0106 & 1.0691 & 1.1109  \\ 
  p-RU-MIDAS HIER && \textbf{0.9646} & \textbf{0.9646} & \textbf{0.9646} & \textbf{0.9646} & \textbf{1.0198} & \textbf{0.9646} &\textbf{ 0.9646}& \textbf{0.9646} & \textbf{0.9646} & \textbf{0.9646} & \textbf{0.9646}  \\ \addlinespace[3pt]
  \hline
\end{tabular}}}
\end{table}

\begin{table}[h]
\centering
\caption{Sensitivity analysis for a rolling window of 10 years. Mean absolute forecast errors, averaged across time, of the HAR, the pooled RU-MIDAS estimated by OLS and the hierarchical estimator, and this for the different macroeconomic variables (columns) and the six forecast horizons. For each horizon, forecast methods in the 75\% Model Confidence Set (MCS) are in bold.}\label{tab:MAFE_MCS_rw10}
\makebox[\textwidth][c]{
\resizebox{1\textwidth}{!}{
\begin{tabular}{llccccccccccc}
  \hline
    Model &\multicolumn{12}{c}{Low-frequency variables} \\ \addlinespace[3pt]
& None & Cpi& Csent & Epui& Fedfunds & Houst& Indpro & M2 & Oil & Termsp & Unrate & Vix \\ 
\hline
 \rowcolor{lightgray} \multicolumn{13}{c}{Horizon $h=1$} \\
  \hline  \addlinespace[3pt]
HAR & \textbf{0.5160} \\ 
  p-RU-MIDAS OLS && \textbf{0.5161} & \textbf{0.5161} & \textbf{0.5156} & \textbf{0.5169} & \textbf{0.5154} & \textbf{0.5162} & \textbf{0.5173} & \textbf{0.5174} & \textbf{0.5166} & \textbf{0.5166} & \textbf{0.5156}  \\ 
  p-RU-MIDAS HIER && \textbf{0.5190} & \textbf{0.5190} & \textbf{0.5190} & \textbf{0.5190} & \textbf{0.5190} &\textbf{ 0.5190} & \textbf{0.5190} & \textbf{0.5190} & \textbf{0.5190} & \textbf{0.5190} &\textbf{ 0.5191} \\ 
  \addlinespace[3pt]
  \hline
  \rowcolor{lightgray} \multicolumn{13}{c}{Horizon $h=5$} \\
  \hline \addlinespace[3pt]
HAR & \textbf{0.6537} \\ 
  p-RU-MIDAS OLS && \textbf{0.6559} & 0.6600 & \textbf{0.6567} & \textbf{0.6583} & \textbf{0.6533} & \textbf{0.6565} & 0.6593 & 0.6591 & \textbf{0.6564} & \textbf{0.6581} & \textbf{0.6546} \\ 
  p-RU-MIDAS HIER && 0.6617 & 0.6617 & 0.6617 & 0.6617 & 0.6617 & 0.6617 & 0.6617 & 0.6617 & 0.6617 & 0.6617 & 0.6617  \\ 
  \hline
 \rowcolor{lightgray} \multicolumn{13}{c}{Horizon $h=20$} \\
  \hline
 HAR & \textbf{0.7733}  \\ 
  p-RU-MIDAS OLS && \textbf{0.7751} & \textbf{0.7812 }& \textbf{0.7698} & \textbf{0.7739} & \textbf{0.7714} & \textbf{0.7774} & \textbf{0.7746} &\textbf{ 0.7736 }& \textbf{0.7762} & \textbf{0.7712} & \textbf{0.7717 } \\ 
  p-RU-MIDAS HIER && 0.7877 & 0.7877 & 0.7878 & 0.7877 & 0.7878 & 0.7870 & 0.7877 & 0.7877 & 0.7877 & 0.7878 & 0.7884  \\ \addlinespace[3pt]
    \hline
   \rowcolor{lightgray} \multicolumn{13}{c}{Horizon $h=40$} \\
\hline \addlinespace[3pt]
HAR & \textbf{0.8403}\\ 
  p-RU-MIDAS OLS && \textbf{0.8395} & 0.8482 & \textbf{0.8285 }& \textbf{0.8408} & \textbf{0.8413} & \textbf{0.8441} & 0.8449& \textbf{0.8457} & \textbf{0.8421} & \textbf{0.8389} & \textbf{0.8374}  \\ 
  p-RU-MIDAS HIER && \textbf{0.8456} & \textbf{0.8452} & \textbf{0.8458} & \textbf{0.8452} &\textbf{ 0.8468} & \textbf{0.8478} & \textbf{0.8452} & \textbf{0.8458} & \textbf{0.8452} & \textbf{0.8478 }& \textbf{0.8478}  \\ \addlinespace[3pt]
   \hline
  \rowcolor{lightgray} \multicolumn{13}{c}{Horizon $h=60$} \\ 
  \hline \addlinespace[3pt]
 HAR & \textbf{0.8867} \\ 
  p-RU-MIDAS OLS && \textbf{0.8856} & 0.8944 & \textbf{0.8738} & \textbf{0.8831} & \textbf{0.8917} & \textbf{0.8917} & \textbf{0.8858} & \textbf{0.8900} & \textbf{0.8822 }& \textbf{0.8846 }& \textbf{0.8819 } \\ 
  p-RU-MIDAS HIER && \textbf{0.8926} & \textbf{0.8926} & \textbf{0.8925} & \textbf{0.8925} & \textbf{0.8925} & \textbf{0.8925} & \textbf{0.8925} & \textbf{0.8926} & \textbf{0.8926} & \textbf{0.8926} & \textbf{0.8924}  \\ \addlinespace[3pt]
   \hline
 \rowcolor{lightgray} \multicolumn{13}{c}{Horizon $h=120$} \\
 \hline \addlinespace[3pt]
 HAR & \textbf{0.9355} \\ 
  p-RU-MIDAS OLS && \textbf{0.9238} & \textbf{0.9372} & \textbf{0.8929} & \textbf{0.9318} & 0.9554 & \textbf{0.9437} & \textbf{0.9286} & \textbf{0.9377} & \textbf{0.9179} & \textbf{0.9356} & \textbf{0.9392}  \\ 
  p-RU-MIDAS HIER && 0.9514 & 0.9514 & 0.9576 & \textbf{0.9443} & \textbf{0.9448} & 0.9514 & 0.9514 & 0.9514 & 0.9514 & 0.9514 & 0.9515  \\ \addlinespace[3pt]
   \hline
\end{tabular}}}
\end{table}

\begin{table}[t]
\centering
\caption{
Sensitivity analysis on the tuning parameter selection. Mean absolute forecast errors, averaged across time, of the pooled RU-MIDAS estimated by the hierarchical estimator with tuning parameter selection by BIC (p-RU-MIDAS HIER BIC) and time series cross-validation (p-RU-MIDAS HIER TSCV), and this for the different macroeconomic variables (columns) and the six forecast horizons and a rolling window of 5 years.}\label{tab:MAFE_MCS_sensitivity_tscv} 
\makebox[\textwidth][c]{
\resizebox{1\textwidth}{!}{
\begin{tabular}{llcccccccccc}
  \hline
    Model &\multicolumn{11}{c}{Low-frequency variables} \\ \addlinespace[3pt]
& Cpi& Csent & Epui& Fedfunds & Houst& Indpro & M2 & Oil & Termsp & Unrate & Vix  \\ 
\hline
 \rowcolor{lightgray} \multicolumn{12}{c}{Horizon $h=1$} \\
  \hline
   \addlinespace[3pt]
  p-RU-MIDAS HIER BIC &  0.4894  &  0.4894 &  0.4894  &  0.4894  &  0.4894  &  0.4894  &  0.4894  &  0.4894  &  0.4894  &  0.4894  &  0.4894 \\
    p-RU-MIDAS HIER TSCV & 0.4907 & 0.4877 & 0.4918 & 0.4957 & 0.4978 & 0.4893 & 0.4905 & 0.4891 & 0.4910 & 0.4905 & 0.4951 \\
    \addlinespace[3pt]
   \hline
  \rowcolor{lightgray} \multicolumn{12}{c}{Horizon $h=5$} \\
   \hline  \addlinespace[3pt]
   p-RU-MIDAS HIER BIC & 0.6210 & 0.6210 & 0.6211 & 0.6210 & 0.6211 & 0.6210 & 0.6210 & 0.6210 & 0.6210 & 0.6210 & 0.6217 \\
  p-RU-MIDAS HIER TSCV & 0.6189 & 0.6195 & 0.6157 & 0.6242 & 0.6222 & 0.6162 & 0.6195 & 0.6180 & 0.6199 & 0.6178 & 0.6235\\
  \addlinespace[3pt]
  \hline
    \rowcolor{lightgray} \multicolumn{12}{c}{Horizon $h=20$} \\
  \hline  \addlinespace[3pt]
  p-RU-MIDAS HIER BIC&  0.7526  &  0.7526  &  0.7526  &  0.7526  &  0.7677  &  0.7526  &  0.7526  &  0.7526  &  0.7526  & 0.7575 &  0.7527  \\   
   p-RU-MIDAS HIER TSCV& 0.7548 & 0.7507 & 0.7447 & 0.7604 & 0.7552 & 0.7494 & 0.7554 & 0.7520 & 0.7498 & 0.7559 & 0.7574  \\
   \addlinespace[2pt]\hline
 \rowcolor{lightgray} \multicolumn{12}{c}{Horizon $h=40$} \\
  \hline  \addlinespace[3pt]
  p-RU-MIDAS HIER BIC&  0.8259  &  0.8271  &  0.8260  &  0.8271  &  0.8528  &  0.8270  &  0.8259  &  0.8271  &  0.8271  &  0.8259  &  0.8237  \\
   p-RU-MIDAS HIER TSCV& 0.8249 & 0.8276 & 0.8168 & 0.8156 & 0.8357 & 0.8204 & 0.8297 & 0.8223 & 0.8216 & 0.8254 & 0.8328\\
   \addlinespace[2pt] \hline
   \rowcolor{lightgray} \multicolumn{12}{c}{Horizon $h=60$} \\
\hline  \addlinespace[3pt]
  p-RU-MIDAS HIER BIC&  0.8534  &  0.8534  &  0.8534  &  0.8534  &  0.9013  &  0.8533  &  0.8534  &  0.8534  &  0.8534  &  0.8533  &  0.8533  \\
  p-RU-MIDAS HIER TSCV & 0.8631 & 0.8548 & 0.8569 & 0.8504 & 0.8903 & 0.8675 & 0.8635 & 0.8564 & 0.8603 & 0.8651 & 0.8880 \\
   \addlinespace[2pt]\hline
  \rowcolor{lightgray} \multicolumn{12}{c}{Horizon $h=120$} \\
  \hline  \addlinespace[3pt]
  p-RU-MIDAS HIER BIC&  0.9726  &  0.9726  &  0.9758  &  0.9669  &  1.0814  &  0.9727  &  0.9726  &  0.9726  &  0.9748  &  0.9831  &  0.9756  \\
   p-RU-MIDAS HIER TSCV & 0.9770 & 0.9814 & 0.9697 & 0.9282 & 1.0640 & 1.0154 & 0.9931 & 0.9826 & 0.9688 & 1.0098 & 1.0662 \\
   \addlinespace[2pt]\hline
\end{tabular}}}
\end{table}

\begin{table}[t]
\centering
\caption{
Sensitivity analysis on the post-lasso estimation. Mean absolute forecast errors, averaged across time, of the pooled RU-MIDAS estimated by the hierarchical estimator with (p-RU-MIDAS HIER PL) and without (p-RU-MIDAS HIER no-PL) a post-lasso step, and this for the different macroeconomic variables (columns) and the six forecast horizons and a rolling window of 5 years.}\label{tab:MAFE_MCS_sensitivity_postlasso} 
\makebox[\textwidth][c]{
\resizebox{1\textwidth}{!}{
\begin{tabular}{llcccccccccc}
  \hline
    Model &\multicolumn{11}{c}{Low-frequency variables} \\ \addlinespace[3pt]
& Cpi& Csent & Epui& Fedfunds & Houst& Indpro & M2 & Oil & Termsp & Unrate & Vix  \\ 
\hline
 \rowcolor{lightgray} \multicolumn{12}{c}{Horizon $h=1$} \\
  \hline
   \addlinespace[3pt]
  p-RU-MIDAS HIER PL &  0.4894  &  0.4894 &  0.4894  &  0.4894  &  0.4894  &  0.4894  &  0.4894  &  0.4894  &  0.4894  &  0.4894  &  0.4894 \\ 
    p-RU-MIDAS HIER no-PL & 0.4926 & 0.4926 & 0.4926 & 0.4926 & 0.4926 & 0.4928 & 0.4926 & 0.4926 & 0.4926 & 0.4925 & 0.4926 \\
    \addlinespace[3pt]
   \hline
  \rowcolor{lightgray} \multicolumn{12}{c}{Horizon $h=5$} \\
   \hline  \addlinespace[3pt]
   p-RU-MIDAS HIER PL & 0.6210 & 0.6210 & 0.6211 & 0.6210 & 0.6211 & 0.6210 & 0.6210 & 0.6210 & 0.6210 & 0.6210 & 0.6217 \\
  p-RU-MIDAS HIER no-PL & 0.6272 & 0.6262 & 0.6286 & 0.6274 & 0.6287 & 0.6271 & 0.6276 & 0.6263 & 0.6278 & 0.6272 & 0.6290 \\
  \addlinespace[3pt]
  \hline
    \rowcolor{lightgray} \multicolumn{12}{c}{Horizon $h=20$} \\
  \hline  \addlinespace[3pt]
  p-RU-MIDAS HIER PL&  0.7526  &  0.7526  &  0.7526  &  0.7526  &  0.7677  &  0.7526  &  0.7526  &  0.7526  &  0.7526  & 0.7575 &  0.7527  \\
   p-RU-MIDAS HIER no-PL& 0.7600 & 0.7606 & 0.7606 & 0.7611 & 0.7759 & 0.7658 & 0.7608 & 0.7602 & 0.7615 & 0.7699 & 0.7644 \\
   \addlinespace[2pt]\hline
 \rowcolor{lightgray} \multicolumn{12}{c}{Horizon $h=40$} \\
  \hline  \addlinespace[3pt]
  p-RU-MIDAS HIER PL&  0.8259  &  0.8271  &  0.8260  &  0.8271  &  0.8528  &  0.8270  &  0.8259  &  0.8271  &  0.8271  &  0.8259  &  0.8237  \\
   p-RU-MIDAS HIER no-PL& 0.8405 & 0.8389 & 0.8414 & 0.8412 & 0.8584 & 0.8420 & 0.8396 & 0.8417 & 0.8419 & 0.8401 & 0.8437 \\
   \addlinespace[2pt] \hline
   \rowcolor{lightgray} \multicolumn{12}{c}{Horizon $h=60$} \\
\hline  \addlinespace[3pt]
  p-RU-MIDAS HIER PL&  0.8534  &  0.8534  &  0.8534  &  0.8534  &  0.9013  &  0.8533  &  0.8534  &  0.8534  &  0.8534  &  0.8533  &  0.8533  \\
  p-RU-MIDAS HIER no-PL& 0.8667 & 0.8663 & 0.8659 & 0.8668 & 0.8983 & 0.8668 & 0.8665 & 0.8662 & 0.8662 & 0.8777 & 0.8689 \\
   \addlinespace[2pt]\hline
  \rowcolor{lightgray} \multicolumn{12}{c}{Horizon $h=120$} \\
  \hline  \addlinespace[3pt]
  p-RU-MIDAS HIER PL&  0.9726  &  0.9726  &  0.9758  &  0.9669  &  1.0814  &  0.9727  &  0.9726  &  0.9726  &  0.9748  &  0.9831  &  0.9756  \\
   p-RU-MIDAS HIER no-PL& 0.9902 & 0.9857 & 0.9873 & 0.9773 & 1.0799 & 0.9832 & 0.9848 & 0.9902 & 0.9900 & 1.0099 & 1.0241 \\
   \addlinespace[2pt]\hline
\end{tabular}}}
\end{table}

\begin{table}[ht]
\centering
\caption{Sensitivity analysis on the high-frequency lag structures. Mean absolute forecast errors, averaged across time, of the HAR, the pooled RU-MIDAS with a dwm-lag structure estimated by OLS (HAR+macro OLS), the hierarchical estimator (HAR+macro HIER), and a hybrid version of the hierarchical estimator where the dwm-lag structure is left unpenalized (HAR+macro HYBRID), and this for the different macroeconomic variables (columns), the six forecast horizons and a rolling window of 5 years.}\label{tab:MAFE_MCS_dwm} 
\makebox[\textwidth][c]{
\resizebox{1\textwidth}{!}{
\begin{tabular}{llccccccccccc}
  \hline 
    Model &\multicolumn{12}{c}{Low-frequency variables} \\
& None& Cpi& Csent & Epui& Fedfunds & Houst& Indpro & M2 & Oil & Termsp & Unrate & Vix  \\ 
\hline 
 \rowcolor{lightgray} \multicolumn{13}{c}{Horizon $h=1$} \\
  \hline \addlinespace[3pt]
  HAR &  0.4866   \\ 
HAR+macro OLS & & 0.4894  &  0.4884  &  0.4900  & 0.4961 &  0.4892  & 0.4913 & 0.4920 & 0.4911 & 0.4915 & 0.4901 & 0.4909  \\ 
  HAR+macro HIER & & 0.4870  &  0.4869  &  0.4871  &  0.4871  &  0.4870  &  0.4869  &  0.4869  &  0.4870  &  0.4869  &  0.4869  &  0.4869   \\ 
   HAR+macro HYBRID && 0.4866 & 0.4866 & 0.4868 & 0.4866 & 0.4866 & 0.4866 & 0.4866 & 0.4866 & 0.4865 & 0.4865 & 0.4865  \\  \addlinespace[3pt]
   
  \hline 
  \rowcolor{lightgray} \multicolumn{13}{c}{Horizon $h=5$} \\
  \hline  \addlinespace[3pt]
  HAR &  0.6097   \\ 
HAR+macro OLS & & 0.6142  & 0.6147 &  0.6131  &  0.6218  &  0.6138  &  0.6136  & 0.6166 & 0.6148 &  0.6144  &  0.6114  &  0.6165  \\ 
  HAR+macro HIER & & 0.6102  &  0.6102  &  0.6107  &  0.6112  &  0.6105  &  0.6100  &  0.6102  &  0.6102  &  0.6101  &  0.6102  &  0.6102   \\ 
  HAR+macro HYBRID && 0.6096 & 0.6096 & 0.6092 & 0.6102 & 0.6108 & 0.6092 & 0.6094 & 0.6095 & 0.6093 & 0.6089 & 0.6090 \\ \addlinespace[3pt]

  \hline 
 \rowcolor{lightgray} \multicolumn{13}{c}{Horizon $h=20$} \\
  \hline  \addlinespace[3pt]
  HAR &  0.7398   \\ 
HAR+macro OLS & &0.7472 &  0.7475  &  0.7377  &  0.7544  &  0.7581 &  0.7446  &  0.7477  &  0.7429 &  0.7471  &  0.7412  &  0.7489  \\ 
  HAR+macro HIER & & 0.7407  &  0.7416  &  0.7407  &  0.7407  &  0.7507  &  0.7408  &  0.7407  &  0.7407  &  0.7404  &  0.7454  &  0.7407   \\ 
 HAR+macro HYBRID &&  0.7404 & 0.7402 & 0.7396 & 0.7400 & 0.7511 & 0.7406 & 0.7404 & 0.7398 & 0.7390 & 0.7471 & 0.7401\\   \addlinespace[3pt]
    \hline
   \rowcolor{lightgray} \multicolumn{13}{c}{Horizon $h=40$} \\
\hline  \addlinespace[3pt]
HAR &  0.8131   \\ 
 HAR+macro OLS & & 0.8172  & 0.8222 &  0.8105  &  0.8183 &  0.8463  &  0.8182  &  0.8270  &  0.8178 &  0.8221  &  0.8192  &  0.8317   \\ 
  HAR+macro HIER & &0.8167 & 0.8166 & 0.8167 & 0.8167 &  0.8361  & 0.8175 & 0.8166 & 0.8167 & 0.8170 & 0.8167 & 0.8161  \\ 
  HAR+macro HYBRID&&  0.8134 & 0.8133 & 0.8131 & 0.8137 & 0.8332 & 0.8134 & 0.8135 & 0.8128 & 0.8131 & 0.8136 & 0.8139 \\ \addlinespace[3pt]
   \hline
   
  \rowcolor{lightgray} \multicolumn{13}{c}{Horizon $h=60$} \\
  \hline  \addlinespace[3pt]
  HAR &  0.8487   \\ 
HAR+macro OLS & & 0.8529  &  0.8561  &  0.8436  &  0.8425  &  0.9017  &  0.8583  &  0.8621  &  0.8547  &  0.8495  &  0.8571 &  0.8840   \\ 
  HAR+macro HIER & & 0.8509  &  0.8521  &  0.8515  &  0.8509  &  0.8817  &  0.8509  &  0.8507  &  0.8524  &  0.8505  &  0.8537  &  0.8512   \\ 
  HAR+macro HYBRID&& 0.8491 & 0.8485 & 0.8487 & 0.8499 & 0.8810 & 0.8486 & 0.8494 & 0.8490 & 0.8487 & 0.8516 & 0.8489 \\
   \addlinespace[3pt]
    \hline
 \rowcolor{lightgray} \multicolumn{13}{c}{Horizon $h=120$} \\
 \hline  \addlinespace[3pt]
 HAR & 0.9850  \\ 
HAR+macro OLS & &0.9785 & 0.9854 & 0.9694 & 0.9121  & 1.0851 & 1.0127 & 0.9943 & 0.9848 & 0.9516 & 1.0015 & 1.0778 \\ 
  HAR+macro HIER & &0.9884 & 0.9883 & 0.9769 & 0.9497 & 1.0902 & 0.9886 & 0.9885 & 0.9883 & 0.9817 & 1.0009 & 1.0126  \\ 
  HAR+macro HYBRID&& 0.9854 & 0.9852 & 0.9739 & 0.9430 & 1.0895 & 0.9853 & 0.9851 & 0.9844 & 0.9780 & 0.9991 & 1.0094  \\
 \addlinespace[3pt]
   \hline
\end{tabular}}}
\end{table}

\end{appendices}
\end{document}